\documentclass[12pt,english,aps,prl,superscriptaddress,showpacs,lengthcheck,longbibliography]{revtex4-1}
\usepackage[T1]{fontenc}
\usepackage[latin9]{inputenc}
\usepackage{float}
\usepackage{amsmath}
\usepackage{amssymb}
\usepackage{graphicx}
\usepackage{esint}

\makeatletter

\providecommand{\tabularnewline}{\\}

\usepackage{babel}

\makeatother

\usepackage{babel}
\begin{document}

\title{Event-Ready Bell Test Using Entangled Atoms Simultaneously Closing
Detection and Locality Loopholes}

\author{Wenjamin Rosenfeld}
\email[corresponding author: ]{W.R@lmu.de}

\affiliation{Fakultät für Physik, Ludwig-Maximilians-Universität München, D-80799
München, Germany}

\affiliation{Max-Planck Institut für Quantenoptik, D-85748 Garching, Germany}

\author{Daniel Burchardt}

\affiliation{Fakultät für Physik, Ludwig-Maximilians-Universität München, D-80799
München, Germany}

\author{Robert Garthoff}

\affiliation{Fakultät für Physik, Ludwig-Maximilians-Universität München, D-80799
München, Germany}

\author{Kai Redeker}

\affiliation{Fakultät für Physik, Ludwig-Maximilians-Universität München, D-80799
München, Germany}

\author{Norbert Ortegel}

\affiliation{Fakultät für Physik, Ludwig-Maximilians-Universität München, D-80799
München, Germany}

\author{Markus Rau}

\affiliation{Fakultät für Physik, Ludwig-Maximilians-Universität München, D-80799
München, Germany}

\author{Harald Weinfurter}

\affiliation{Fakultät für Physik, Ludwig-Maximilians-Universität München, D-80799
München, Germany}

\affiliation{Max-Planck Institut für Quantenoptik, D-85748 Garching, Germany}
\begin{abstract}
An experimental test of Bell's inequality allows ruling out any local-realistic
description of nature by measuring correlations between distant systems.
While such tests are conceptually simple, there are strict requirements
concerning the detection efficiency of the involved measurements,
as well as the enforcement of spacelike separation between the measurement
events. Only very recently could both loopholes be closed simultaneously.
Here we present a statistically significant, event-ready Bell test
based on combining heralded entanglement of atoms separated by $398\,\mathrm{m}$
with fast and efficient measurements of the atomic spin states closing
essential loopholes. We obtain a violation with $S=2.221\pm0.033$
(compared to the maximal value of $2$ achievable with models based
on local hidden variables) which allows us to refute the hypothesis
of local-realism with a significance level $P<2.57\cdot10^{-9}$. 
\end{abstract}

\pacs{03.65.Ud, 32.80.Qk}

\maketitle
Back in 1935 Einstein, Podolsky and Rosen (EPR) pointed at inconsistencies
in quantum mechanics, if one requires that a physical theory has to
be realistic and local \cite{Einstein1935}. In such theories any
signal, influence, or interaction propagates at most at the speed
of light (\emph{locality}), and one can assign properties to quantum
systems \emph{before} a measurement (\emph{realism}). To achieve the
latter, they left open the possibility to complement quantum mechanics
with, nowadays called, \emph{local hidden variables} (LHV). Starting
from the EPR example on analyzing measurement results of two independent
observers, John Bell showed that the prediction of QM for certain
measurement scenarios differ from the prediction of all local, realistic
theories \cite{Bell1964}. With this he directly provided a prescription
for how to evaluate the validity of the EPR claims and of any LHV
theory in an experiment.

However, there are stringent requirements on an experimental test,
as LHVs give a theory an amazing flexibility to account for observed
results. In spite of the many experiments started soon after Bell's
discovery (e.g. \cite{Freedman1972,Aspect1982a}), which (almost)
all agreed well with QM, they all relied on assumptions on the observers
or the observed systems, thus opening loopholes to the LHV theories
under test (for reviews see, e.g., \cite{Clauser1978,Tittel2001,Larsson2014}).

One loophole, the \emph{locality loophole}, concerns the independence
of the observers, which only can be warranted if the whole measurement
processes of the two observers are spacelike separated. This was achieved
by Weihs et al. \cite{Weihs1998}, where the whole measurement, starting
from the choice of a random number up to the appearance of the classical
voltage signal of a single photon detection was outside the light
cone of the other measurement. However, as detection of single photons
was notoriously inefficient those days, one had to assume fair sampling,
i.e. that the registered photon pairs had been a representative sample
of all pairs - thus leaving open the so called \emph{detection loophole}.
This was closed for the first time in an experiment using trapped,
entangled ions \cite{Rowe2001}, which, however, were separated only
by few micrometers - leaving the locality loophole open. Since then
the goal was to close both in a single experiment, leading to key
developments such as the first observations of atom-photon entanglement
\cite{Blinov2004,Volz2006} and atom-atom entanglement over larger
distances \cite{Moehring2007, *Ritter2012, Hofmann2012}. Recently,
based on electron spins of separated nitrogen-vacancy (NV) centers
\cite{Jelezko2006} the first experimental test of Bell's theorem
without the locality and detection loophole was performed \cite{Hensen2015}.
With the development of efficient photon pair sources \cite{Fedrizzi2007, *Trojek2008}
and highly efficient single photon detectors \cite{Lita2008} two
tests succeeded also with entangled photon pairs \cite{Giustina2015,Shalm2015}.

\begin{figure*}
\begin{centering}
\includegraphics{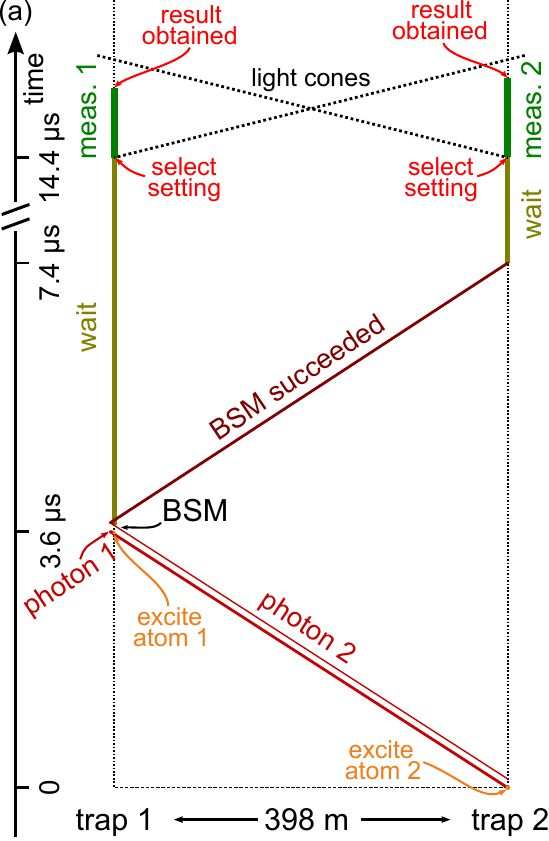}\hfill{}\includegraphics[scale=0.4]{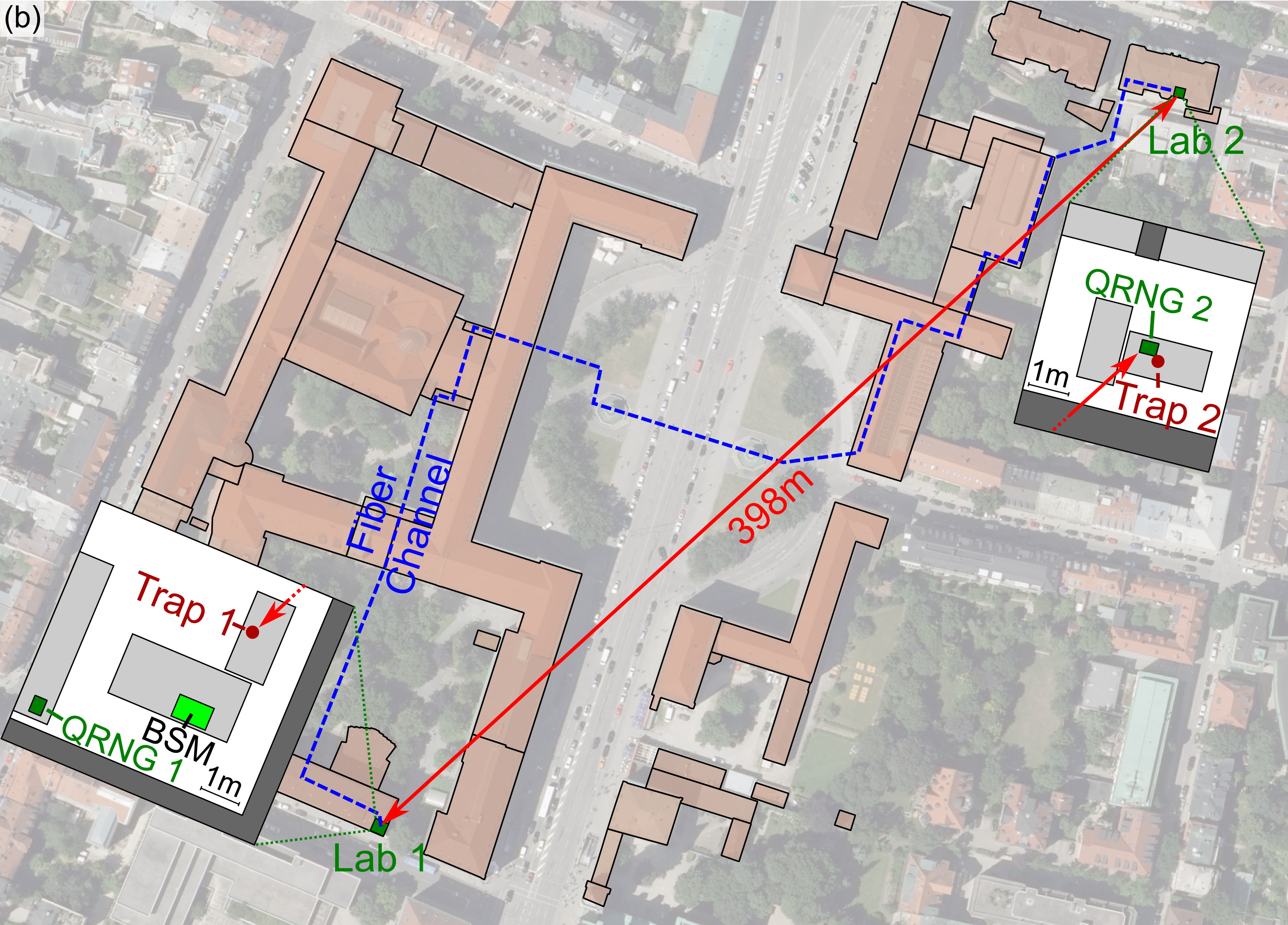} 
\par\end{centering}
\caption{\label{fig:1-SpaceTimeDiagram}(a) Space-time diagram of the experiment.
The two observers (trap~1 and trap~2) are separated by $398\,\mathrm{m}$
with the BSM setup being located close to trap~1. Single photons
and all communication signals are transmitted via optical fibers (lengths
vary around $700\,\mathrm{m}$) laid in cable ducts connecting the
two stations. Sending a photon from trap~2 to the BSM takes roughly
$3.6\,\mathrm{\mu s}$ (photons from both traps arrive within a window
of $120\,\mathrm{ns}$ represented by two lines for earliest and latest
emission). Another $3.7\,\mathrm{\mu s}$ are needed for communicating
the success of the BSM back to trap~2. The state measurements (including
random choice of the measurement direction) are performed such that
a result is obtained outside of the light cone of the other side.
(b) Overview of the experimental location on the main campus of LMU.
Trap~1 is located in the basement of the faculty of physics and trap~2
in the basement of the department of economics. Map data were provided
by \cite{vermessungsamt}. BSM: Bell state measurement, QRNG: quantum
random number generator.}
\end{figure*}

Here we describe the evaluation of LHV theories using entangled neutral
atoms closing both the locality and the detection loophole in a single
experiment. Based on atom-photon entanglement, entanglement swapping
\cite{Zukowski1993} allowed to prepare in a heralded manner entangled
spin states of two atoms separated by a distance of $398\,\mathrm{m}$,
well suited for an event-ready test. For an event-ready test no fair
sampling assumption has to be made \cite{Bell1988,Zukowski1993}.
There a measurement result is reported every time the heralding signal
confirming the successful distribution of entanglement to the observers
was obtained and thus no detection loophole is opened at all. Any
inefficiencies or inaccuracies in the atomic state detection then
only influence the degree of achievable correlations. The locality
loophole is closed by employing fast and efficient measurements of
the atomic spin states at a sufficient distance together with fast
quantum random number generators (QRNG) for selection of the measurement
basis. We employed state-dependent ionization for highly efficient
atomic state analysis and with a total observation time of about a
microsecond also the spacelike separation could be warranted. Well-defined
hypothesis tests with samples of $10000$ observations clearly indicate
that LHV theories do not allow a correct description of nature.

We consider the simplest situation of an event-ready Bell test, where
two separate observers are told - according to a heralding signal
- to report the result of two-outcome measurements $A$, $B$ $\in\{\uparrow,\downarrow\}$
performed on each side (an example are measurements on spin-$\frac{1}{2}$
particles). For a test of local realism the two observers choose their
measurement directions from two possibilities $a\in\{\alpha,\alpha'\}$
and $b\in\{\beta,\beta'\}$ and afterwards compare their results.
For this situation Clauser, Horne, Shimony, and Holt (CHSH) put Bell's
inequality in an experimentally friendly form \cite{Clauser1969}:
\begin{equation}
S=\left|\left\langle \sigma_{\alpha}\sigma_{\beta}\right\rangle +\left\langle \sigma_{\alpha}\sigma_{\beta'}\right\rangle \right|+\left|\left\langle \sigma_{\alpha'}\sigma_{\beta}\right\rangle -\left\langle \sigma_{\alpha'}\sigma_{\beta'}\right\rangle \right|\leq2,\label{eq:CHSH}
\end{equation}
with correlators $\left\langle \sigma_{a}\sigma_{b}\right\rangle =\frac{1}{N_{a,b}}(N_{a,b}^{\uparrow\uparrow}+N_{a,b}^{\downarrow\downarrow}-N_{a,b}^{\uparrow\downarrow}-N_{a,b}^{\downarrow\uparrow})$.
Here $N_{a,b}^{A,B}$ denote the number of events with the respective
outcomes $A$, $B$ for measurement directions $a$, $b$ and $N_{a,b}$
is the total number of events of the respective measurement setting.
Quantum mechanics predicts a violation of this inequality when measurements
are performed on maximally entangled states $\left|\Psi^{\pm}\right\rangle =\frac{1}{\sqrt{2}}\left(\left|\uparrow\right\rangle \left|\downarrow\right\rangle \pm\left|\downarrow\right\rangle \left|\uparrow\right\rangle \right)$
with certain measurement settings, e.g., $\alpha=0^{\circ}$, $\alpha'=90^{\circ}$,
$\beta=-45^{\circ}$, $\beta'=45^{\circ}$. Angles $\alpha,\beta$
are defined here in the spin space.

\begin{figure*}
\begin{centering}
\includegraphics{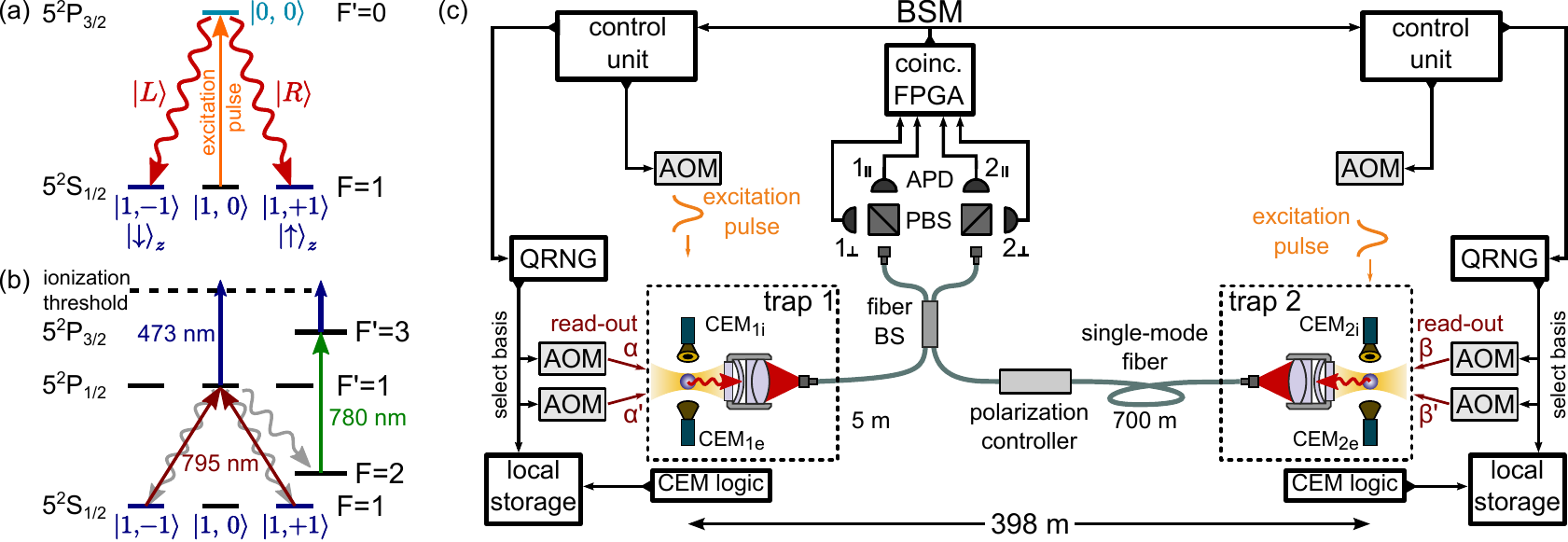} 
\par\end{centering}
\caption{\label{fig:2-Level-Experiment-Scheme} (a) Scheme of the atomic levels
involved in the entanglement between the spin state of the atom (subspace
$5^{2}S_{1/2}$, $F=1$, $\left|m_{F}=\pm1\right\rangle $) and polarization
of the photon (left- and right-circular, $\left|L\right\rangle $,
$\left|R\right\rangle $, respectively). Entanglement is generated
in the spontaneous decay of the $5^{2}P_{3/2}$, $F'=0$ state after
optical excitation. (b) Scheme of the atomic state measurement. A
selected superposition of the spin states is excited to the $5^{2}P_{1/2}$,
$F'=1$ level depending on the polarization of a $795\,\mathrm{nm}$
laser pulse and is ionized with a $473\,\mathrm{nm}$ laser. The atom
can spontaneously decay to the $5^{2}S_{1/2}$, $F=1$ or $F=2$ levels
during this procedure (gray wavy arrows). While decays into the $F=1$
level can reduce the fidelity of the measurement process, population
in the $F=2$ level is excited with an additional $780\,\mathrm{nm}$
laser and ionized as well. (c) Schematic of the experimental setup.
In each trap spin-polarization entanglement is generated between the
atom and a single photon which is guided to the BSM via a single-mode
fiber. Polarization stability in the $700\,\mathrm{m}$ fiber connecting
trap~2 and the BSM is ensured by automatic compensation \cite{Rosenfeld2008}
performed every $5\,\mathrm{min}$ using reference light and a polarization
controller. The photons are overlapped on a fiber beam splitter (BS),
their coincident detection heralds entanglement of the atomic spins.
Local measurements are performed on the atomic spins according to
settings selected by quantum random number generators (QRNGs). AOM:
acousto-optic modulator, APD: avalanche photo diode, CEM: channel
electron multiplier, FPGA: field programmable gate array, PBS: polarizing
beam-splitter.}
\end{figure*}

In our case the two observer stations are independently operated setups
(trap~1 and trap~2) that are equipped with their own laser and control
systems. Their separation of $398\,\mathrm{m}$ (Fig. \ref{fig:1-SpaceTimeDiagram})
makes $1328\,\mathrm{ns}$ available to warrant spacelike separation
of the measurements. On each side we store a single $^{87}\mathrm{Rb}$
atom in an optical dipole trap. The employed internal spin states
($\left|\uparrow\right\rangle _{z}$ and $\left|\downarrow\right\rangle _{z}$)
are the Zeeman states $\left|m_{F}=+1\right\rangle $ and $\left|m_{F}=-1\right\rangle $
of the ground level $5^{2}S_{1/2}$, $F=1$ (Fig. \ref{fig:2-Level-Experiment-Scheme}(a)).
Entanglement of the atoms is generated by first entangling the spin
of each atom with the polarization of a single emitted photon \cite{Volz2006}.
The photons are guided to an interferometric Bell state measurement
(BSM) setup (Fig. \ref{fig:2-Level-Experiment-Scheme}), located close
to trap~1. It consists of a fiber beam splitter (BS) followed by
polarizing beam splitters (PBS) in each of the output ports, where
detection of photons is performed by four avalanche photodiodes (APDs).
This setup allows to distinguish two maximally entangled photon states.
Thereby a two-photon coincidence in particular detector combinations
(see Sec. I.B of the Supplemental Material \cite{supplement}, which
includes Refs. \cite{Glauber2007,Loudon1997,LEcuyer2007,Zhang2013,McDiarmid1989,Bednorz2015,Adenier2015})
heralds the projection of the atoms onto one of the states $\left|\Psi^{\pm}\right\rangle =\frac{1}{\sqrt{2}}\left(\left|\uparrow\right\rangle _{x}\left|\downarrow\right\rangle _{x}\pm\left|\downarrow\right\rangle _{x}\left|\uparrow\right\rangle _{x}\right)$
\cite{Hofmann2012}, where $\left|\uparrow\right\rangle _{x}=\frac{1}{\sqrt{2}}\left(\left|\uparrow\right\rangle _{z}+\left|\downarrow\right\rangle _{z}\right)$
and $\left|\downarrow\right\rangle _{x}=\frac{i}{\sqrt{2}}\left(\left|\uparrow\right\rangle _{z}-\left|\downarrow\right\rangle _{z}\right)$.

The experimental sequence (see Supplemental Material \cite{supplement}
Sec. I.B for further details) starts after two atoms are loaded into
the traps. Photons emitted by the atoms are coupled into optical fibers.
The efficiencies for detecting a single photon in the BSM arrangement
after excitation in trap~1 or trap~2 are $\eta_{1}=1.65\times10^{-3}$
and $\eta_{2}=0.85\times10^{-3}$ (the latter also includes the transmission
loss of photons ($\lambda=780\,\mathrm{nm}$) in the $700\,\mathrm{m}$
fiber of approximately $50\%$). This results in an overall probability
to obtain a heralding signal in the BSM of $0.7\times10^{-6}$. If
no signal is obtained the excitation sequence of the atoms is repeated.
Including times necessary for transmission of signals as well as to
prepare and to cool the atoms, the average rate of excitation attempts
is $5.2\times10^{4}\,\mathrm{s^{-1}}$. Depending on the loading rate
of the traps this results in about $1$ - $2$ heralding events per
minute. The atom excitation procedures are synchronized to $<1\,\mathrm{ns}$
(Supplemental Material \cite{supplement} Sec. I.A) such that the
emitted photons entangled with the respective atoms have, at the BSM
setup, a temporal overlap close to unity \cite{Hofmann2012}.

After a successful BSM signals are sent to both observers where they
trigger the switching to atomic state measurement. An additional waiting
time has to be introduced due to dephasing and rephasing of atomic
states in strongly focused dipole traps. There, longitudinal field
components lead to an inhomogeneous light polarization which results
in a state- and position-dependent AC Stark shift. Due to the antisymmetry
of the polarization distribution this accumulated phase is compensated
after one transverse oscillation \cite{RosenfeldPhD2008,*Thompson2013,*TBP01}.
To obtain simultaneous rephasing the radial trap frequencies are chosen
for an oscillation period $\frac{2\pi}{\omega_{r}}$ of $11.2\,\mu s$
and $14.5\,\mathrm{\mu s}$ for trap~1 and trap~2, respectively,
by setting the trap depths. The measurement procedure starts with
selecting the analysis direction according to the output of a fast
quantum random number generator. As a further development of \cite{Fuerst2010}
these QRNGs have minimal bias (typ. less than $10^{-5}$) without
any postprocessing \cite{Abellan2015}. The random bit in trap~1
(trap~2) determining the direction $\alpha/\alpha'$ ($\beta/\beta'$)
is provided on request and has no measurable correlation to bits generated
earlier than $80\,\mathrm{ns}$ before, see the Supplemental Material
\cite{supplement} Sec. II for details. In the sense of independence
to previous information, we thus consider this moment before the request
as the starting time of the measurement.

For the analysis of the atomic state a state-selective ionization
is employed where the measurement direction $\gamma\in\{\alpha,\alpha',\beta,\beta'\}$
is determined by the polarization of a readout laser at $795\,\mathrm{nm}$
exciting the atom to the $5^{2}P_{1/2}$, $F'=1$ level from where
it is ionized by an additional laser at $473\,\mathrm{nm}$ (Fig.
\ref{fig:2-Level-Experiment-Scheme}(b)). In particular, we ionize
the state $\left|\uparrow\right\rangle _{\gamma}=\sin(\gamma/2)\left|\uparrow\right\rangle _{x}-\cos(\gamma/2)\left|\downarrow\right\rangle _{x}$
using linear polarization at an angle $\gamma/2$ relative to the
horizontal. The state $\left|\downarrow\right\rangle _{\gamma}=\cos(\gamma/2)\left|\uparrow\right\rangle _{x}+\sin(\gamma/2)\left|\downarrow\right\rangle _{x}$
remains unaffected. The resulting $^{87}\mathrm{Rb}^{+}$-ion and
electron are accelerated by an electric field to two channel electron
multipliers (CEMs) placed in $8\,\mathrm{mm}$ distance from the trapping
region. The ionization fragments are detected with high efficiencies
$\eta_{i}=0.90..0.94$ (ions), $\eta_{e}=0.75..0.90$ (electrons),
the efficiencies are slightly different for the two labs and also
vary between different measurement runs. We assign detection of at
least one of the fragments to the atomic state $\left|\uparrow\right\rangle _{\gamma}$,
providing a total detection efficiency of $\geq0.98$ \cite{Henkel2010,OrtegelPhD2016},
while detection of no fragment is assigned to the state $\left|\downarrow\right\rangle _{\gamma}$.
Note that in the event-ready scheme an imperfect detection efficiency
does only affect the fidelity of the measurement process.

In order to perform a fast selection of the measurement direction
we switch on one of two polarized readout laser beams with an acousto-optical
modulator (AOM) (Fig. \ref{fig:2-Level-Experiment-Scheme}). The latency
time from the output of the random bit of the QRNG until the readout
pulse reaches the atom is $217\,(204)\,\mathrm{ns}$. Optimizing the
measurement fidelity we accept ions arriving at the detectors up to
$570\,(725)\,\mathrm{ns}$ after the beginning of the ionization process.
The different times for the two traps result from different acceleration
fields and, consequently, different times of flight of the ions. Together
with the avalanche transition time within the CEMs and the latency
of the processing electronics of $80\,(84)\,\mathrm{ns}$, the total
time until the result appears as a digital pulse at the output is
$947\,(1093)\,\mathrm{ns}$ after the starting time of the measurement.
We consider this signal being perfectly clonable and, thus, representing
a definite classical entity with a value existing independent of observation.
It is recorded together with the respective random bit (at trap~1
also with the result of the BSM) in a local storage unit.

We performed several measurement runs in the time period between November
2015 and June 2016. After a first clear violation with $300$ events
could be observed on Nov. 27, 2015 (see \cite{supplement} Sec. VI.A),
the stability of the setup was improved allowing for long-term measurements.
For testing the hypothesis that our experimental results can be described
by a LHV theory, a well-defined experimental procedure was established
to avoid expectation bias \cite{Jeng2006}. For that purpose all relevant
details were fixed before the start of each run. These include the
number of events to be collected, the analysis procedure, as well
as scheduled maintenance to be performed, see the Supplemental Material
\cite{supplement} Sec. IV. We chose $5000$ events for each prepared
atomic state to achieve an appropriate level of significance, evaluation
according to Eq. (\ref{eq:CHSH}) and maintenance every $24$ hours.
We present two runs fulfilling these criteria in the following.

\begin{figure}
\begin{centering}
\includegraphics[scale=0.5]{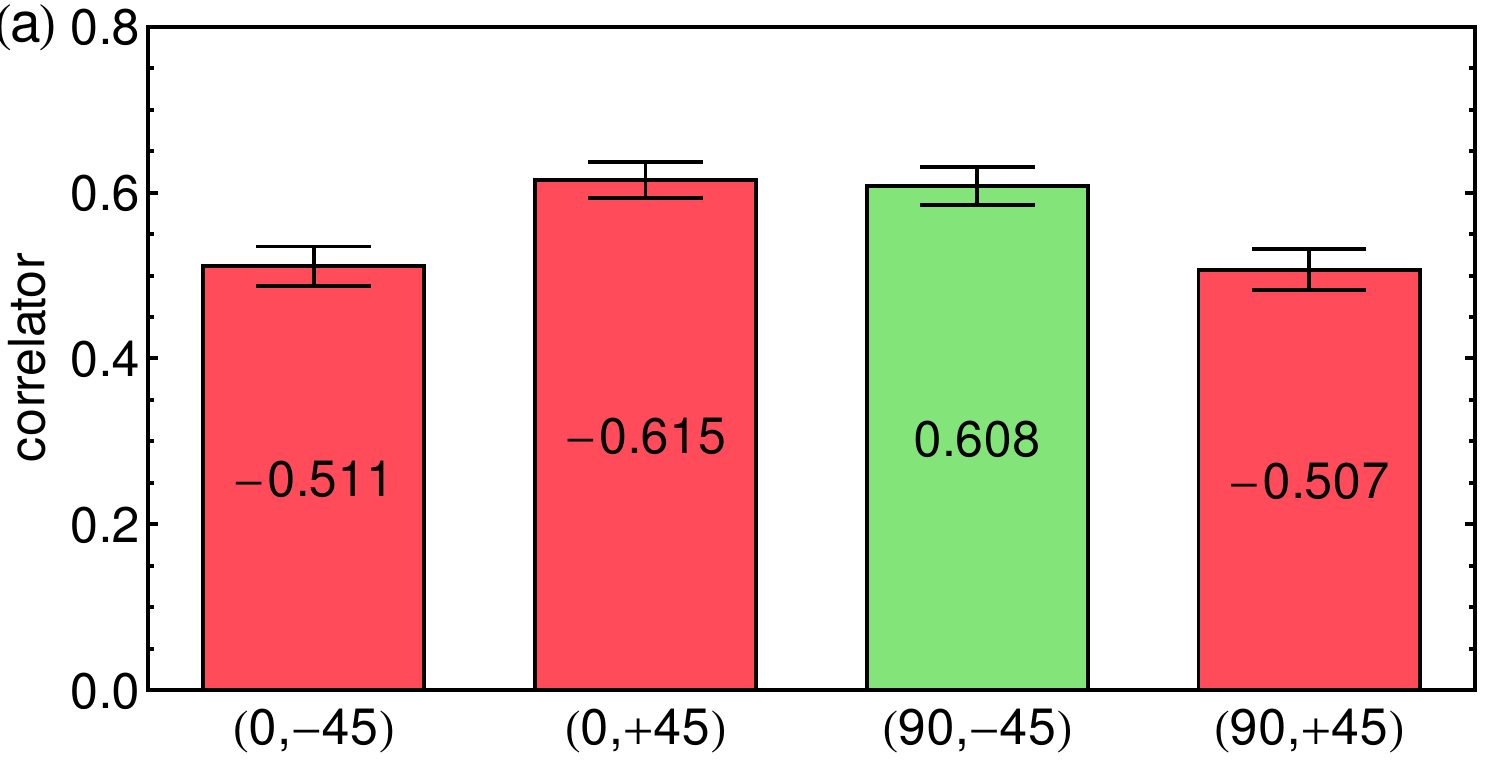} 
\par\end{centering}
\begin{centering}
\includegraphics[scale=0.5]{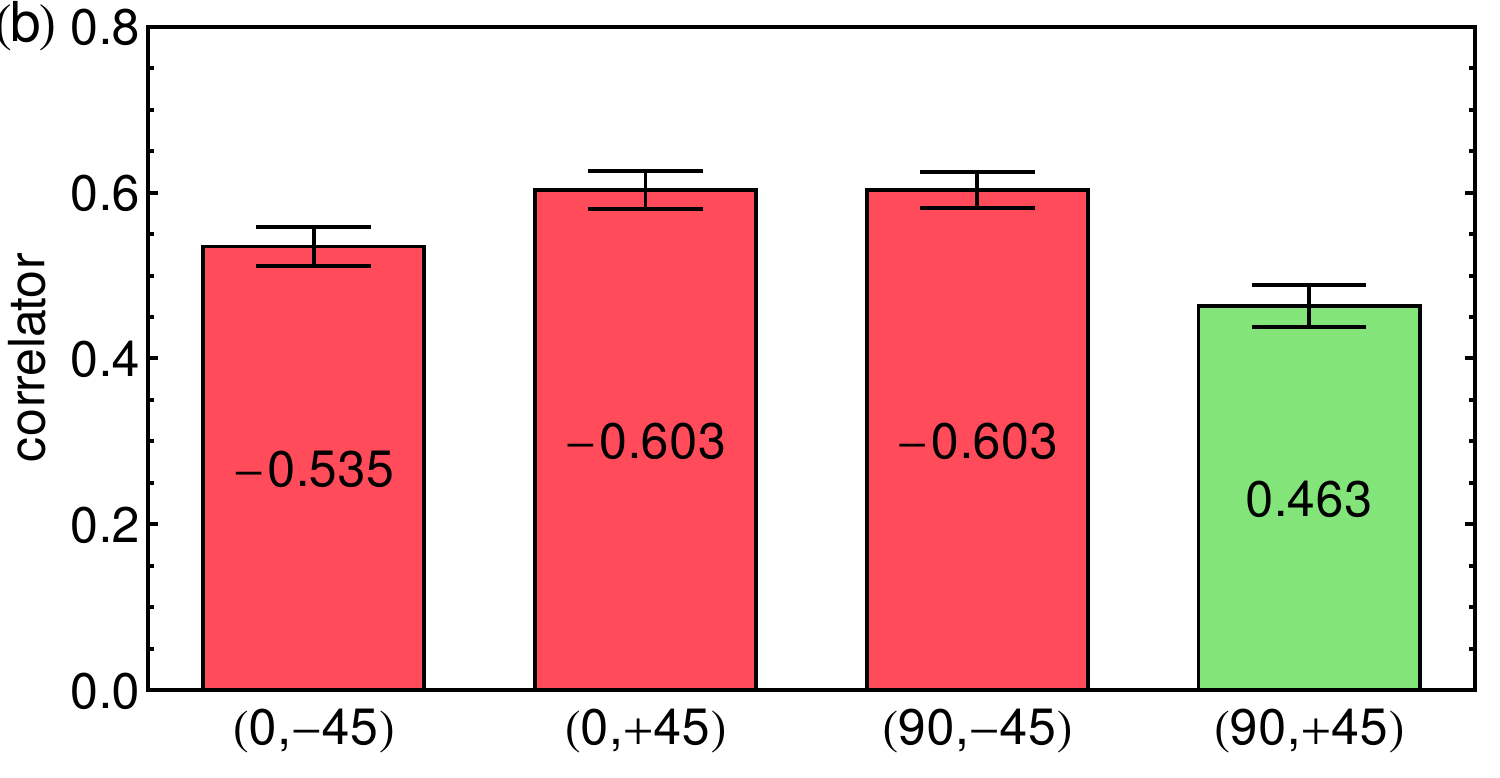} 
\par\end{centering}
\caption{\label{fig:Correlators01}Measured correlators $\left\langle \sigma_{a}\sigma_{b}\right\rangle $
for the run started on Apr.~15, 2016 for the atom-atom state $\left|\Psi^{-}\right\rangle $
(a) and $\left|\Psi^{+}\right\rangle $ (b). Displayed errors are
equal to one standard deviation.}
\end{figure}

For the measurement run started on Apr. 15, 2016 the obtained correlations
are shown in Fig. \ref{fig:Correlators01}. For the $5000$ events
for each of the two atom-atom states collected during $4$ days, the
resulting $S$-parameters of $2.240\pm0.047$ ($\left|\Psi^{-}\right\rangle $)
and $2.204\pm0.047$ ($\left|\Psi^{+}\right\rangle $) show a violation
of the LHV limit by $5.1$ and $4.3$ standard deviations, respectively.
By combining the events for the two atomic states we obtain $S=2.221\pm0.033$
corresponding to a violation by $6.7$ standard deviations.

In order to determine the impact of these results for ruling out LHV
theories we use the null hypothesis that the experiment is governed
by LHV. Under this assumption one can estimate the probability of
obtaining a certain violation of Bell's inequality or a more extreme
one, which is called the P-value. Within the hypothesis one can also
allow for potential \emph{memory} effects \cite{Barrett2002}, where
the history of the experiment may influence the probabilities of outcomes.
We use two different models for calculating upper bounds for the P-value:
the martingale approach \cite{Gill2003} ($P_{m}$) and the game formalism
\cite{Brunner2014, *Elkouss2016} ($P_{g}$), for details see \cite{supplement}
Sec. III. For the combined data of the measurement above we obtain
$P_{m}=2.57\cdot10^{-9}$ and $P_{g}=1.74\cdot10^{-10}$.

Explicit data for the above run, for the first violation in 2015,
as well as of further runs are documented in the Supplemental Material
\cite{supplement} Sec. VI. Especially, we want to point at the run
started on June 14, 2016. The start of it was made public via the
Twitter account \emph{@munichbellexp} and simultaneously at a conference
\cite{Vaexjoe2016}. The results of each of the events, coming in
at a rate of about $1/\mathrm{min}$, were directly communicated to
a central server \url{http://bellexp.quantum.physik.uni-muenchen.de},
which made all the data available together with the momentary evaluation.
In this public Bell test, due to the lower rate of trapping single
atoms the $2\times5000$ events were collected during a time of $10$
days, resulting in $S=2.134\pm0.048$ ($\left|\Psi^{-}\right\rangle $)
and $S=2.057\pm0.048$ ($\left|\Psi^{+}\right\rangle $). The violations
of $2.8$ and $1.2$ standard deviations result in P-values for the
combined data of $P_{m}=0.0267$ and $P_{g}=2.82\cdot10^{-3}$. It
should be noted, that with the modest event rate the effect of counting
statistics on the momentary value of the S-parameter became clearly
visible to a wide audience. The complete data are available for download
from the server.

Finally, we consider a further frequently mentioned loophole - the
\emph{free-will} (or\emph{ freedom of choice}) loophole \cite{Bell1985}
targeting the independence of choice of the analysis directions from
the hidden variables and vice versa \cite{Larsson2014}. Contrary
to experiments with photon pairs \cite{Giustina2015,Shalm2015}, event-ready
tests using entanglement swapping do not have a typical moment where
the LHVs would have been defined \cite{Branciard2010}. If we assume
that the LHVs are defined at the time of the BSM, in our experiment
taking place $10.7\,\mu s$ before the choice of the local analysis
directions, they are clearly not influenced by the latter. Yet, contrary,
the random settings are determined within the light cone of the BSM
and independence has to be assumed here. This was accounted for in
\cite{Hensen2015,Giustina2015,Shalm2015} where generation of the
random numbers is considered being outside of the light cone of the
entanglement generation (allowing to exclude influences within one
trial of the experiment up to a few nanoseconds for the photon experiments
\cite{Giustina2015,Shalm2015} or $690\,\mathrm{ns}$ for the experiment
using NV-centers \cite{Hensen2015}).

However, in the analysis of all experiments (including the present
one) there is still the implicit assumption that the dependence of
the random numbers generated for the $n$-th observation event on
processes or events of any kind in their backward light cone is strongly
limited \footnote{There are well-developed descriptions accounting for a possible dependence
of the random settings on the history of the experiment \cite{Kofler2016,Elkouss2016}
(see Supplemental Material \cite{supplement} Eq. (10)). Yet, this
dependence is effectively set to a very small value, only depending
on technical noise evaluated within a physical model (\cite{Abellan2015}
and Supplemental Material \cite{supplement} Sec. II), and thus one
effectively assumes independence of the generation process from the
history.}(e.g., dependence on previous settings and outcomes of the experiment).
Effectively, while one allows memory and by this dependence on the
history for the LHV model determining the measurement outcomes, one
does not allow memory for the (quantum) systems observed in the QRNGs
to determine the settings. To avoid such assumptions - and the corresponding
loopholes - and to warrant true independence of the random settings
also in view of memory attributed to \emph{all} quantum systems one
should produce random numbers outside the light cones of \emph{all}
other events of the Bell test. Spacelike separated extraterrestrial
sources of randomness are required and have to be developed to ensure
this \cite{Gallicchio2014, *Handsteiner2017}.

In this Letter we described a highly reliable event-ready Bell test,
showing in several attempts a clear violation of a Bell inequality.
With violations of more than $6$ standard deviations obtained in
a run with $10000$ events the probability that this actual result
could be described by local hidden variables is at most $P_{m}=2.57\cdot10^{-9}$.
Taking all data accumulated during a time period of $7$ months with
over $55000$ events (without any postselection) decreases this value
to $P_{m}=1.02\cdot10^{-16}$. On the fundamental side, further reducing
the number of assumptions on the independence of the randomness generation
makes the development of methods for employing extraterrestrial sources
highly desirable. From the point of view of applications, where the
requirements for the random setting choice are different, our essentially
loophole-free Bell test forms a promising platform for device-independent
secure communication. The methods and results achieved here pave the
way for new developments of quantum information and for future quantum
repeater networks.
\begin{acknowledgments}
We gratefully acknowledge the help of all the people who contributed
to the development of the experiment over the last $15$ years, especially
C. Kurtsiefer, M. Weber, J. Volz, F. Henkel, M. Krug, J. Hofmann,
and we thank M. Zukowski for fruitful discussions. We acknowledge
funding by the German Federal Ministry of Education and Research via
the projects QuOReP and Q.com-Q, by the German-Israeli Foundation
Project I-282-303.9-2013, and by the EU ERC Project QOLAPS.
\end{acknowledgments}

\bibliography{belltest_refs}

\newpage{}

\onecolumngrid
\setcounter{equation}{0}
\setcounter{figure}{0}
\setcounter{table}{0}

\makeatletter
\renewcommand{\theequation}{S\arabic{equation}}
\renewcommand{\thefigure}{S\arabic{figure}}
\renewcommand{\thetable}{S\arabic{table}}

\newpage{}

\part*{Supplemental material}

\newpage{}

\section{Experimental Details}

\subsection{Synchronization and control of the experiment}

\begin{figure}[b]
\begin{centering}
\includegraphics{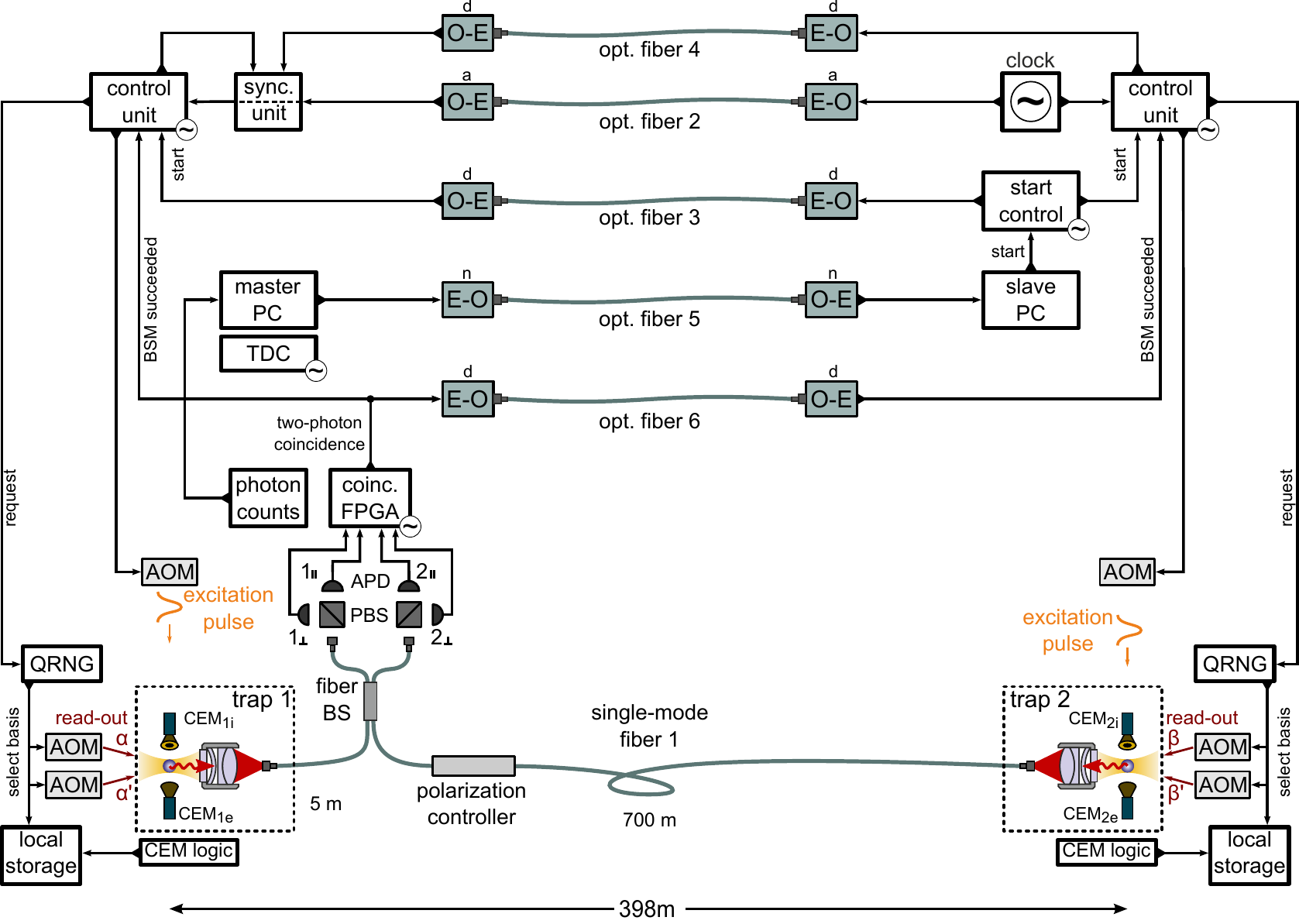}
\par\end{centering}
\caption{\label{fig:Synchronization-control}Synchronization and control of
the experiment. Some signals are omitted for simplicity. Devices marked
with a \textcircled{\textasciitilde{}} symbol are synchronized to
the common $100\,\mathrm{MHz}$ clock. AOM: acousto-optic modulator,
APD: avalanche photo diode, BS: beam splitter, CEM: channel electron
multiplier, EO/OE: electro-optic/opto-electric converters for analog
(a), digital (d) or network (n) signals, FPGA: field programmable
gate array, PBS: polarizing beam-splitter, QRNG: quantum random number
generator. }
\end{figure}

The experiment consists of two independent atom traps connected by
a $700\,\mathrm{m}$ long fiber link for photons emitted by trapped
atoms as well as for synchronization and communication between the
two setups. All communication between the two sides is done optically
at telecom wavelength (Fig. \ref{fig:Synchronization-control}, fibers
2-6). Depending on the specific requirements the signals are transmitted
via analog~(a), digital (d) or network (n) electro-optical converters.
Each trap is controlled locally by a PC and a custom-built pattern
generator that is used as a control unit (CU). The PC next to trap~1
acts as master. It is capable of controlling the remote PC in lab~2
by sending commands via an optical network connection (fiber 5). It
also continuously analyzes the photon counts collected from both traps
via the 4 avalanche photo diodes (APDs) of the BSM arrangement. The
CUs operate at a clock speed of $50\,\mathrm{MHz}$ and thus a resolution
of $20\,\mathrm{ns}$ with a timing jitter of $<40\,\mathrm{ps}$.
They are capable of switching lasers and logic signals in a programmable
way and also respond to external signals. All timing-critical components
are synchronized to a common $100\,\mathrm{MHz}$ clock located in
lab~2 whose signal is distributed via fiber 2. Synchronization of
the CUs is monitored by a synchronization unit in lab~1 with the
help of an additional signal transmitted via fiber 4 guaranteeing
altogether a low relative jitter of less than $150\,\mathrm{ps}$
rms betwen the two labs. A time-to-digital converter (TDC) with a
time resolution of $80\,\mathrm{ps}$, connected to the master PC,
records all necessary time stamps, e.g., all photons counts and time
markers generated by CUs indicating the different sequences in the
experiment.

\subsection{Experimental sequence}

Loading of the atom traps is controlled by the PCs. Depending on the
level of photon counts integrated within $40\,\mathrm{ms}$, it is
distinguished whether atoms are present in the two traps and loading
operations are initialized accordingly. Loading of both atom traps
takes typically about $2-3\,\mathrm{s}$. 

\begin{figure}
\begin{centering}
\includegraphics{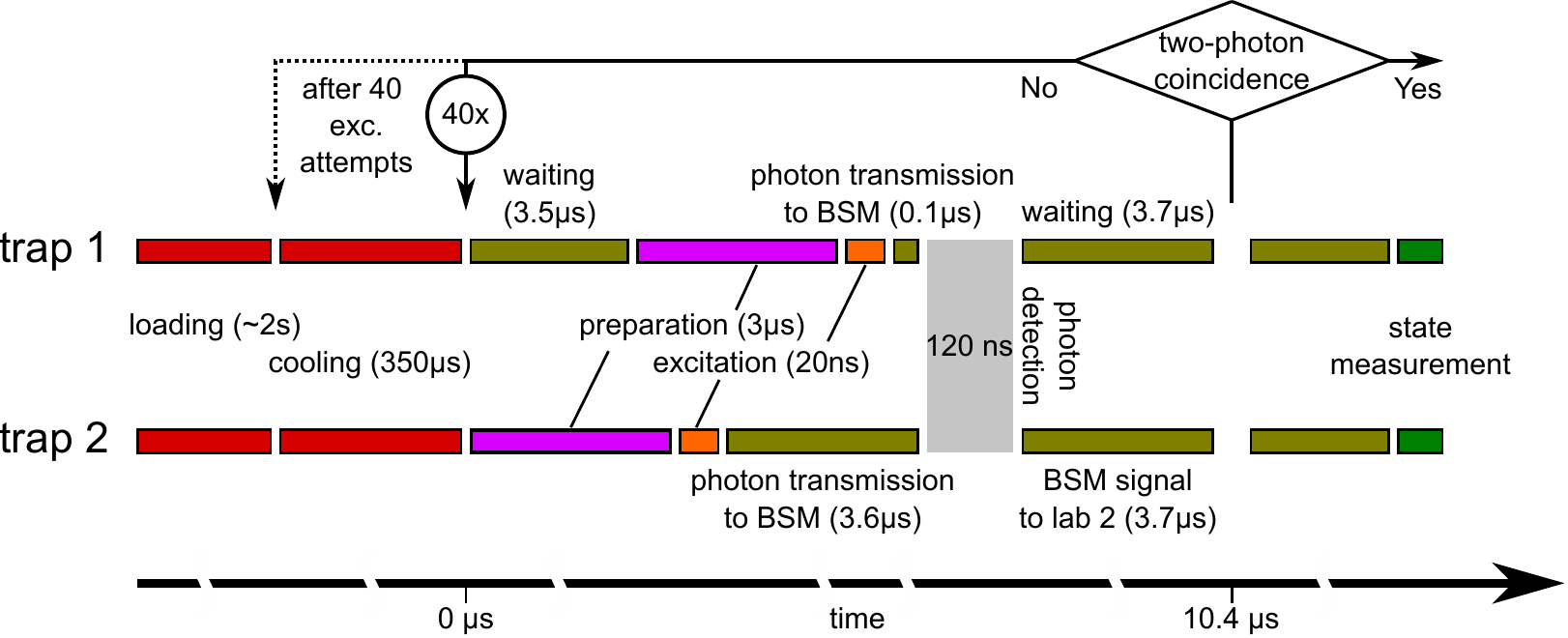}
\par\end{centering}
\caption{\label{fig:Experimental-sequence}Experimental sequence.}
\end{figure}

After both traps are loaded the master PC initiates switching of the
two CUs to the excitation sequence (Fig. \ref{fig:Experimental-sequence}).
The required signal is generated by the start control unit (SCU) in
lab~2 and communicated to lab~1 via fiber~3 (Fig. \ref{fig:Synchronization-control}).
The excitation sequence consists of preparation of the $5^{2}S_{1/2},F=1,m_{F}=0$
by optical pumping followed by excitation to the $5^{2}P{}_{3/2},F'=0,m_{F'}=0$
state. After each excitation there is a waiting time of $7.3\,\mathrm{\mu s}$
needed to transmit the photons from trap~2 to the BSM setup in lab~1
and to transmit a potential two-photon detection signal back to lab~2.
This procedure is performed in repeated bursts that are timed such
that the emitted photons of both trap setups have a temporal overlap
close to unity at the BSM arrangement. A successful BSM is registered
by one of four characteristic two-photon detection events ($1_{\perp}2_{\parallel}$
or $1_{\parallel}2_{\perp}$ for $\left|\Psi^{-}\right\rangle $ and
$1_{\perp}1_{\parallel}$ or $2_{\perp}2_{\parallel}$ for $\left|\Psi^{+}\right\rangle $)
in a time-window of $120\,\mathrm{ns}$ \cite{Hofmann2012}. If one
of those two-photon detections has occurred, signals are sent to both
CUs to switch to the state measurement. Otherwise after $40$ preparation-excitation
cycles the atoms are recooled for $350\,\mathrm{\mu s}$ and the sequence
is restarted giving an average rate of excitation attempts is $5.2\times10^{4}\,\mathrm{s^{-1}}$.
If the master PC registers loss of one of the atoms the corresponding
trap is reloaded.

\subsection{Detailed timing scheme of the atomic state measurement\label{subsec:Detailed-timing-scheme}}

\begin{figure}
\begin{centering}
\includegraphics[scale=0.9]{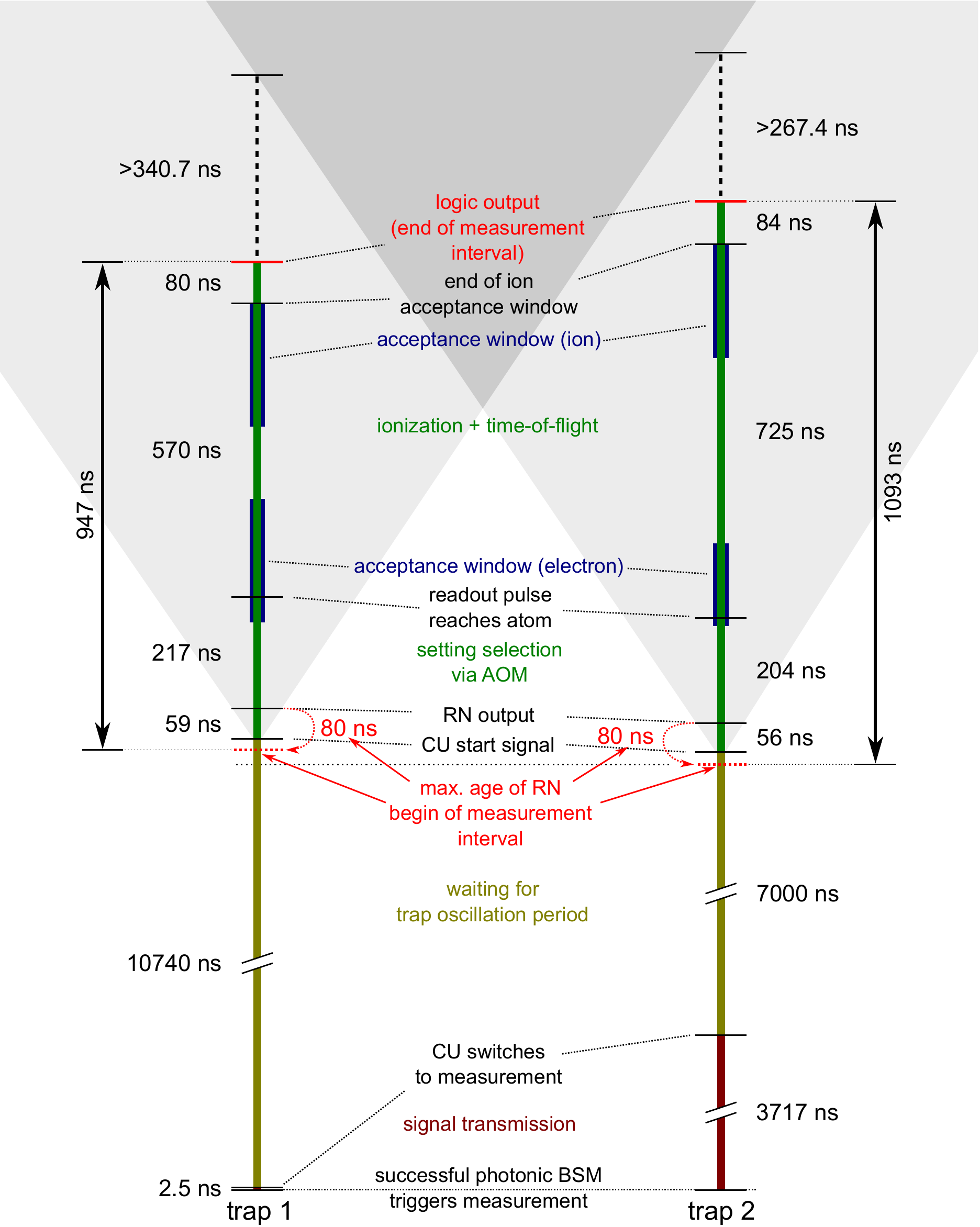}
\par\end{centering}
\caption{\label{fig:Detailed-timing-scheme}Detailed timing scheme of the experimental
sequence after successful BSM.}
\end{figure}

Fig. \ref{fig:Detailed-timing-scheme} shows the measured timings
of all relevant processes of the atomic state measurement after a
two-photon coincidence. The state-measurements have to be space-like
separated which requires fast measurements with precisely known timings.

The common start signal for the state-measurement is generated in
lab~1 by a FPGA registering a two-photon coincidence in the BSM arrangement.
This signal is distributed to both control units~(CUs) locally via
a $50\,\mathrm{cm}$ coaxial cable and via the $700\,\mathrm{m}$
long optical link for the distant location. The propagation time is
measured by the delay of a signal transmitted to the remote location
and back. This yields signal transmission times of $2.5\pm0.2\,\mathrm{ns}$
and $3717\pm7\,\mathrm{ns}$, respectively. Additional waiting times
on both sides are necessary to minimize dephasing of the atomic state
which are chosen such that the state measurement is performed after
a full transverse oscillation period of the atom in the respective
trap. The CU in lab~1 waits $10.74\,\mathrm{\mu s}$ and in lab~2
waits $7.00\,\mathrm{\mu s}$, respectively, before a measurement
start signal is issued. With this the measurements in the two traps
start almost simultaneously (trap~2 starts $28.5\,\mathrm{ns}$ earlier).

First, a random bit is requested determining the measurement direction
and, in our setup, determining which of the two acousto-optic modulators
(AOMs) is activated to switch readout laser beam with the respective
polarization. The switching times of the AOMs, given by the propagation
of the acoustic wave in the crystal, together with propagation delays
in cables, electronic components, and the optical path of the readout
pulse to the position of the atom are measured with fast photo diodes
placed in an equivalent distance to the atom traps. The delays between
the output of a random bit and the beginning of the ionization process
are $217\pm4\,\mathrm{ns}$ and $204\pm4\,\mathrm{ns}$. The time
of flight of electrons towards the channel electron multipliers (CEMs)
was calculated to be $3\,\mathrm{ns}$. Using this, the ionization
time and time of flight of the ions to the CEMs are extracted from
the arrival time histogram of electrons and ions. With these times
at the beginning of measurement run we define two fixed time-windows
for accepting electron and ion detection events chosen to optimize
the signal to noise ratio for better fidelity. In the presented measurement
runs the lengths of these windows were $240\,\mathrm{ns}$ ($160\,\mathrm{ns}$)
for electrons and $240\,\mathrm{ns}$ ($220\,\mathrm{ns}$) for ions
for trap~1 (trap~2). The maximal time for ionization and flight
given by the end of the ion acceptance window is $570\pm3\,\mathrm{ns}$
and $725\pm3\,\mathrm{ns}$ for the two traps, respectively (the uncertainty
results from the rising edge in the electron arrival time histogram
at the beginning of the process). These are different for two traps
as the acceleration voltages between the CEMs differ to avoid high-voltage
breakdowns in trap~2.

The measurement is considered finished when the detection clicks of
the CEMs are transformed to a logic pulse in electronics outside the
vacuum chamber which happens $80\,\mathrm{ns}$ and $84\,\mathrm{ns}$
after the detection. Together with the time of the RN generation/correlation
of $80\,\mathrm{ns}$ this results in the overall measurement time
of $947\pm1\,\mathrm{ns}$ and $1093\pm1\,\mathrm{ns}$, respectively.
Note that this time is known with a high precision as it is based
on signals of the CU (starting signal of the measurement and the end
of the ion acceptance window). The stated uncertainty results mostly
from the additional signal processing units (like signal converters,
etc.). 

\subsection{Criteria of space-like separation}

The distances between the laboratories and in particular between the
QRNGs and atom traps were determined by combining measurements within
the buildings with maps and building outlines provided by the Bavarian
land surveying office \cite{vermessungsamt}. The positions of the
atom trap and the QRNG in each lab were determined with respect to
a reference position at the outer corner of the building in three
dimensions with an accuracy of better than $10\,\mathrm{cm}$. The
accuracy of this measurement guarantees that for each point the real
position is within a volume of $0.5\,\mathrm{m}$ radius around the
measured position. The coordinates of the reference points, the orientations
of the buildings and respective altitudes were extracted from the
provided data \cite{vermessungsamt}. By these means the measured
distances between the QRNGs and atom traps are $398.0\,\mathrm{m}$
(QRNG~2 $\leftrightarrow$ trap~1) and $402.7\,\mathrm{m}$ (QRNG~1
$\leftrightarrow$ trap~2). Assuming maximal errors favoring the
shortest distance we arrive at minimal time for a luminal signal to
reach the other side of $(398.0\,\mathrm{m}-2\cdot0.5\,\mathrm{m})/c=1324.2\,\mathrm{ns}$
and $(402.7\,\mathrm{m}-2\cdot0.5\,\mathrm{m})/c=1339.9\,\mathrm{ns}$,
respectively. 

Taking the values from section \ref{subsec:Detailed-timing-scheme},
the maximal measurement times are $947+1\,\mathrm{ns}=948\,\mathrm{ns}$
and $1093+1\,\mathrm{ns}=1094\,\mathrm{ns}$. The uncertainty of the
signal transmission time (Sec. \ref{subsec:Detailed-timing-scheme})
leads to an uncertainty in the measurement start time for trap~2
of $\pm7\,\mathrm{ns}$. Since the atomic state measurement in trap~2
starts $28.5\,\mathrm{ns}$ earlier, space-like separation is therefore
guaranteed with remaining margins of $1324.2\,\mathrm{ns}-28.5\,\mathrm{ns}-7\,\mathrm{ns}-948\,\mathrm{ns}=340.7\,\mathrm{ns}$
and $1339.9\,\mathrm{ns}+28.5\,\mathrm{ns}-7\,\mathrm{ns}-1094\,\mathrm{ns}=267.4\,\mathrm{ns}$
(Fig. \ref{fig:Detailed-timing-scheme}). By further delaying the
measurement at trap~1 the margin could in principle be made symmetric
at $304.0\,\mathrm{ns}$.

\subsection{Data recording}

For evaluation of the experiment we use two independent ways for data
recording. First, for every successful BSM event, the resulting Bell
state, the requested random bit and the measurement outcomes (CEM
signals) are recorded locally on both sides (``local storage'' in
Fig. \ref{fig:Synchronization-control}). This enables an evaluation
of correlations and of Bell's inequality. Second, the TDC records
time stamps of all photons registered in the BSM arrangement, as well
as of the requested random bits and CEM detection clicks from both
sides. Additional signals indicating the position in the experimental
sequence are stored in this data stream enabling a complete analysis
of the experiment.

\newpage{}

\section{Generation of Random Settings}

For performing a test of Bell's inequality free of the locality loophole
the choices of the measurement directions, in the ideal case, are
perfectly unpredictable. In our experiment these choices are derived
from the outputs of quantum random number generators (QRNGs) which
are unpredictable according to the physical model of the QRNGs. Technical
imperfections can lead to a residual predictability of the output
bit sequences which has to be taken into account in the analysis of
the experimental results (Sec. \ref{sec:Testing-LHV-Theories}). In
particular to derive the P-values one requires the maximal deviation
$\tau$ from perfect unpredictability, which is defined such that
for all output bits $q_{i}$: $\frac{1}{2}-\tau\text{\ensuremath{\leq}}Pr(q_{i}=0)\leq\frac{1}{2}+\tau$.
In the following we describe the function of the employed QRNGs and
estimate their predictability from a model. Furthermore we perform
evaluation of bias, serial correlations and general statistical tests.
Although statistical testing can never certify real randomness, it
is the method of choice for testing the hypothesis of the bit sequence
being random. Moreover, it still gives some information about the
quality of randomness of the bits obtained from the QRNGs and allows
identifying potential artifacts.

\subsection{Random number generators}

The method used for the random number generator (QRNG) \cite{Fuerst2010}
is based on counting the number of photons emerging from a light emitting
diode (LED) source, passing an attenuator and detected by a photo-multiplier
tube (PMT). The analog pulses at the output of the PMT are digitized
by a comparator and counted within time bins of $20\,\mathrm{ns}$.
The parity of the registered photon number finally constitutes the
random bit.

The physical model employs the fact that according to photodetection
theory \cite{Glauber2007,Loudon1997} the detection events from a
broadband light source of constant power are fully uncorrelated on
the timescale of the counting interval. In particular, each photon
is assumed to be registered by the detector at an unpredictable time
which is independent of any previous events in the backward light-cone
as well as of the LHVs in the experiment. This leads to a Poissonian
distribution of the detected photon numbers at the PMT. In our implementation
the extendable dead time of the detector, i.e. its inability to register
pulses arriving in a short succession which are interpreted as one
long pulse, modifies this distribution \cite{Fuerst2010}, Fig.2.
This is analogous to a real-time hardware processing additionally
allowing to avoid bias of the output bits without the need for any
further post-processing \cite{Fuerst2010}. Here we also assume that
the detector itself can not be influenced within the LHV model.

In the experiment the QRNGs are operated at a speed of $50\,\mathrm{Mbps}$
and provide the last generated random bit on request within $8\,\mathrm{ns}$.
Including all hardware latencies, the maximal ``age'' of this bit
is $60\,\mathrm{ns}$ since the emission of the detected, contributing
photon(s). The generators exhibit a modest next neighbor correlation
of typically $<1.5\cdot10^{-5}$ while all evaluated correlations
with higher lag were found to be compatible with zero (see below).
We thus include $20\,\mathrm{ns}$ for generation of the previous
random bit and consider the output to be at most $80\,\mathrm{ns}$
``old''. All bits generated during an experimental run were continuously
recorded for analysis purposes (organized in files of $1\,\mathrm{Gb}$)
\footnote{Data can be made available on request.}. The QRNGs incorporate
continuous stabilization and monitoring of temperature and count rate
to allow for long-term operation.

\subsection{Estimation of predictability}

While the emission and detection of photons are intrinsically random
fundamental processes at the current state of knowledge, there are
further technical parameters which may affect the output bit. Depending
on the model, such parameters may be accessible when the outputs $A$,
$B$ are generated by the observers and thus increase the predictability.
The crucial parameters of the generators used are the average photon
count rate, the threshold level of the comparator digitizing the PMT
output pulses (influencing the extendable dead time \cite{Fuerst2010})
and the temperature of the QRNG devices.

\subsubsection*{Photon count rate}

For a given threshold of the PMT pulse comparator, the bias of the
QRNG is a function of the photon count rate. Thus knowledge of the
count rate allows a certain predictability. Before each measurement
run the QRNGs are operated for a longer period to determine the count
rate for minimal bias. During the measurement run the count rate is
stabilized at this value by a feedback loop controlling the LED current.
Once fixed, the count rate shows only expected statistical fluctuations.
While this already shows the high stability of the involved components,
we have additionally characterized the sensitivity of the bias to
the LED current leading to effects much lower then the error made
when determining the optimal count rate for minimal bias. This leads
to a residual bias of typically between $10^{-6}$ and $10^{-5}$
(see Tab. \ref{tab:Bias}).

\begin{figure}
\begin{centering}
\includegraphics{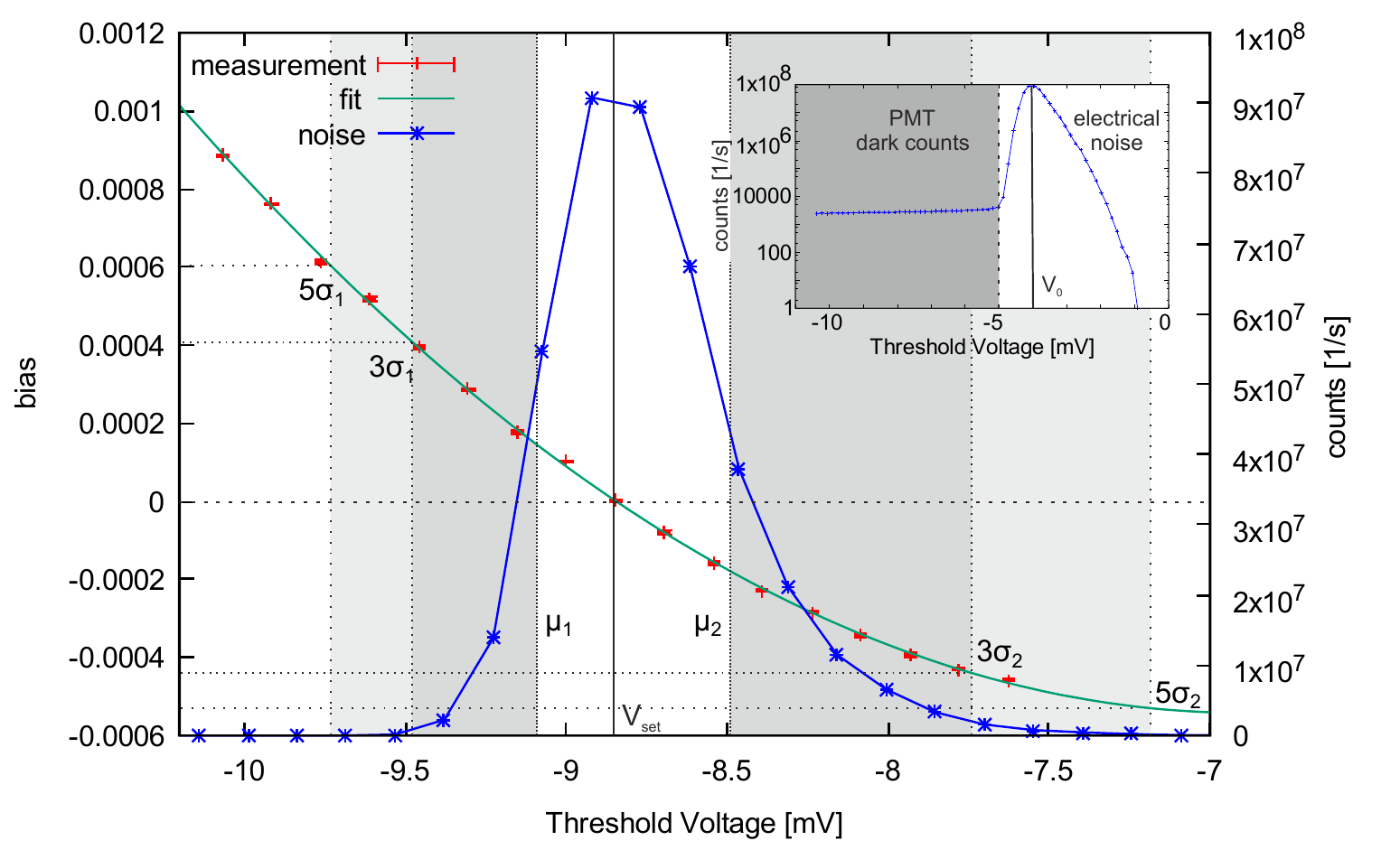}
\par\end{centering}
\centering{}\caption{\label{fig:Predictability}Measured bias (red crosses, approximated
by second order polynomial green fit curve) as a function of the comparator
threshold voltage level for the QRNG operated at typical LED current.
The inset shows the distribution of registered counts as a function
of the threshold voltage for LED switched off revealing the electrical
noise and dark counts of the PMT. The blue stars in the main graph
show the measurement from the inset on a linear scale and additionally
centered around the set value of the threshold voltage $V_{set}$
(thus being equivalent to an integrated histogram of effective fluctuations
of the set value). We approximate the rising and falling slope of
the electrical noise part by error functions of the form $a\cdot(\mathrm{erf}(\frac{\mu-v_{thr}}{\sqrt{2}\sigma})+1)$
from both sides yielding the $3\sigma$ and $5\sigma$ intervals of
amplitude fluctuations. This allows us to determine the maximal predictability
for the corresponding confidence levels.}
\end{figure}

\subsubsection*{Threshold level of the comparator}

Any fluctuations of the threshold level, or complementary any noise
on the signal from the PMT will lead to a time-dependent bias and
predictability. We thus analyze the bias for different threshold levels
at a constant LED current. The result is shown in Fig. \ref{fig:Predictability}(red
crosses). In a second measurement we have determined the distribution
of registered counts with LED being switched off. Here we can observe
two regions - one dominated by the dark counts of the PMT ($<-5\,\mathrm{mV}$)
and one dominated by the electrical noise (Fig. \ref{fig:Predictability},
inset). For further analysis we assume that the measured data is an
integrated histogram of the the noise amplitude distribution. We thus
approximate the rising and the falling slopes of the electrical noise
part (\ref{fig:Predictability}, blue stars) which is centered around
the threshold set voltage by error functions obtaining the mean value
$\mu_{1}=-9.09\,\mathrm{mV}$ and standard deviation $\sigma_{1}=0.13\,\mathrm{mV}$
on the rising side and $\mu_{2}=-8.48\,\mathrm{mV}$ and standard
deviation $\sigma_{2}=0.25\,\mathrm{mV}$ on the falling side. Since
the electrical noise adds to the real PMT pulses, it can be considered
equivalent to noise of the comparator threshold level. Thus the influence
of this noise on the predictability (bias) can be directly derived
from these numbers. For a ``paranoid'' model which grants the LHV-controlled
observers full knowledge about the noise of the QRNG in the distant
lab we can now consider either the average over the time-dependent
predictability or merely the maximal predictability (within a $5\sigma$
confidence interval). In the latter, most paranoid, case we arrive
at a maximal additional predictability of $\tau=6.12\cdot10^{-4}$.

\subsubsection*{Temperature of the QRNG device}

The temperature of the critical part of the QRNG including LED, PMT
and the comparator is actively stabilized to better than $\pm0.15\,\mathrm{^{\circ}C}$.
Still, the residual fluctuations, which may be accessible, can influence
the critical threshold level of the comparator. Here the specified
temperature coefficients of the comparator offset voltage and of the
DAC providing the threshold level are both $10^{-5}\,\mathrm{V/^{\circ}C}$.
This gives together with the bias dependence from above (\ref{fig:Predictability})
an additional predictability of $\tau=6.7\cdot10^{-6}$ (in the case
the effects add up).

\subsubsection*{Resulting predictabilities}

Let us distinguish two models: 
\begin{itemize}
\item a (reasonable) model where the information on the internal parameters
of the QRNG are inaccessible except for temperature and the count
rate (which are both stabilized and monitored with information being
available externally). The resulting deviations of the generated bit
sequence from an ideal random one determine the bias of the sequence.
The maximal observed bias over all measurements (Tab. \ref{tab:Bias})
is $8.74\cdot10^{-6}$, compatible with the expectation. Taking this
value and additionally adding a $2\sigma$ margin to it we can conservatively
estimate a deviation from perfect unpredictability of $\tau_{1}=1.04\cdot10^{-5}$.
\item a (very paranoid) model where also the full information on the internal
noise at the comparator is known and can be used when determining
the measurement result according to such LHV models. This allows for
an additional predictability which might not be visible in any typical
statistical test. Note that exploiting this information for the distant
lab requires extrapolation of the noise behavior for $1.3\,\mathrm{\mu s}$
into the future. With this ability granted, we sum over all above
effects (error in the setting of photon count rate, noise on the threshold
level, temperature dependence) arriving at a $\tau_{2}<6.3\cdot10^{-4}$.
We take this value for calculation of all P-values allowing us to
exclude even such models.
\end{itemize}
We note that the predictabilites, even in the paranoid models, can
be significantly reduced by performing an XOR operation on several
successive output bits thereby combining them into one bit. This operation
can be efficiently performed in hardware. In our experiment the time
budget allows for combining at least $13$ bits, while $\tau$ would
be $<10^{-6}$ already for a combination depth of $2$ bit.

\subsection{Bias}

\begin{figure}
\begin{centering}
\includegraphics{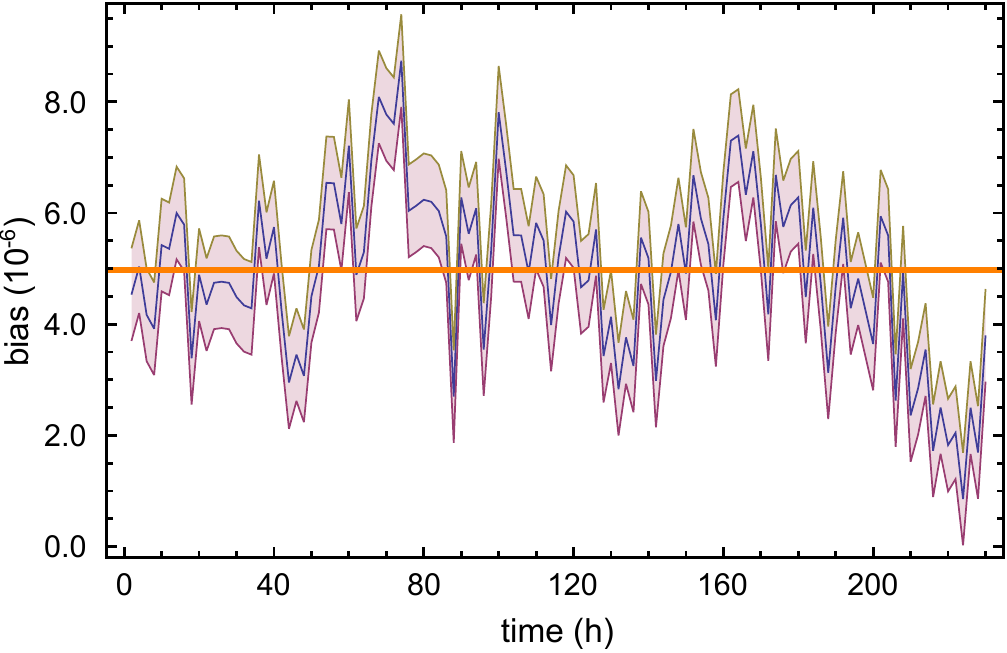}
\par\end{centering}
\caption{\label{fig:Bias-time-evolution}Time evolution of the bias in lab~1
for the measurement run on June 14, 2016. Each bin contains $360\,\mathrm{Gb}$
of data, corresponding to $2\,\mathrm{h}$ of measurement time. The
shaded area represents a $\pm\sigma$ interval. The orange line shows
the average value.}
\end{figure}

We have evaluated the bias $B=\frac{n_{0}}{n}-\frac{1}{2}$ ($n_{0}$
being the number of ones in a sample of $n$ bits) of the generated
bit sequences. Figure \ref{fig:Bias-time-evolution} shows an example
of the observed bias as a function of time for one of the measurement
runs. The bin size is chosen large enough for the statistical noise
to be smaller than the observed value, which still allows observing
possible drifts on a long time scale. Table \ref{tab:Bias} gives
an overview of the data and the maximal observed bias for different
measurement runs.

\begin{table}
\begin{centering}
\begin{tabular}{|c|c|c|c|c|}
\hline 
run & \#bits & $\max(\left|B\right|)_{a}$ & $\max(\left|B\right|)_{b}$ & $\sigma$\tabularnewline
\hline 
\hline 
Nov. 27, 2015 & $2\times10.6\,\mathrm{Tb}$ & $2.66\cdot10^{-6}$ & $4.76\cdot10^{-6}$ & $8.33\cdot10^{-7}$\tabularnewline
\hline 
Apr. 04, 2016 & $2\times8.0\,\mathrm{Tb}$ & $4.04\cdot10^{-6}$ & $2.99\cdot10^{-6}$ & $8.33\cdot10^{-7}$\tabularnewline
\hline 
Apr. 15, 2016 & $2\times16.0\,\mathrm{Tb}$ & $4.33\cdot10^{-6}$ & $3.22\cdot10^{-6}$ & $8.33\cdot10^{-7}$\tabularnewline
\hline 
June 14, 2016 & $2\times41.6\,\mathrm{Tb}$ & $8.74\cdot10^{-6}$ & $7.15\cdot10^{-6}$ & $8.33\cdot10^{-7}$\tabularnewline
\hline 
\end{tabular}
\par\end{centering}
\caption{\label{tab:Bias}Bias for various measurement runs. $\sigma$ corresponds
to the standard deviation in a $360\,\mathrm{Gb}$ bin.}
\end{table}

\subsection{Correlations}

\begin{figure}
\begin{centering}
\includegraphics{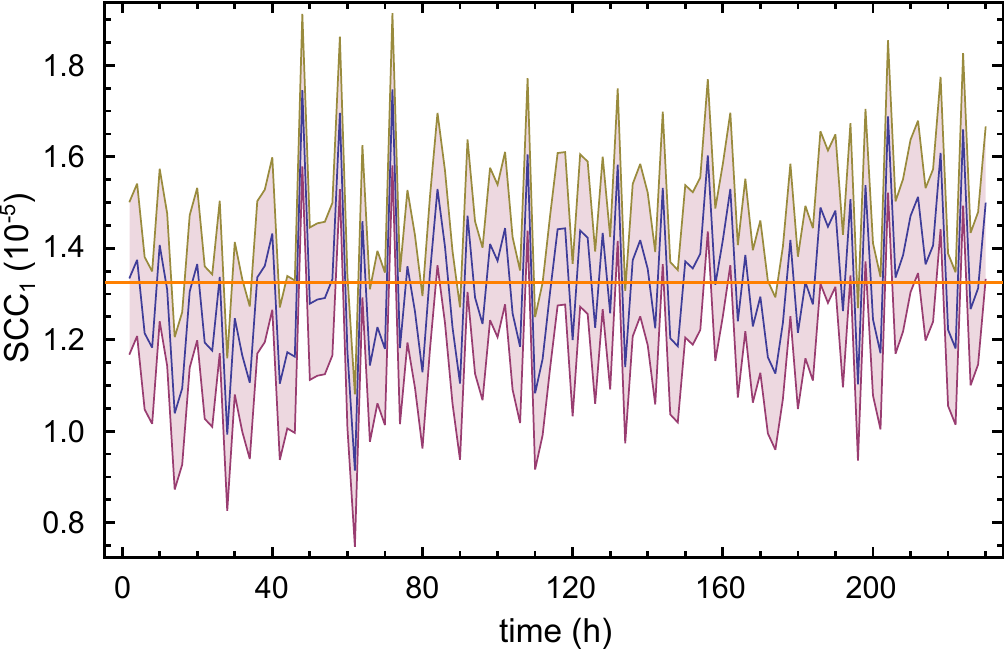}
\par\end{centering}
\caption{\label{fig:SCC-time-evolution}Time evolution of $SCC_{1}$ in lab~1
for the measurement run on June 14, 2016. Each bin contains $360\,\mathrm{Gb}$
of data, corresponding to $2\,\mathrm{h}$ of measurement time. The
shaded area represents a $\pm\sigma$ interval. The orange line shows
the average value of $1.32\cdot10^{-5}$.}
\end{figure}

For each measurement and both random number generators we analyzed
the data of every file for serial correlations (or autocorrelation)
according to 
\[
SCC_{l}=\frac{\sum_{k=1}^{n-l}(q_{k}-\frac{1}{2})(q_{k+l}-\frac{1}{2})}{\sum_{k=1}^{n}(q_{k}-\frac{1}{2})^{2}}
\]
where $q_{k}\in\{0,1\}$ are the bits and $l$ is the lag (we evaluated
correlations up to a lag of $56$). This formula is not corrected
for bias which in our case would lead only to negligible corrections.
All correlations with lag $l>1$ were found to be consistent with
zero. Tab. \ref{tab:Correlations} exemplarily shows the correlations
for one of the measurement runs. We also evaluated the time evolution
of $SCC_{1}$, only a small variation could be observed within $\sim230\,\mathrm{h}$
of measurement time, see Fig. \ref{fig:SCC-time-evolution}.

\begin{table}
\begin{centering}
\begin{tabular}{|c|c||c|c||c|c||c|c||c|c||c|c|}
\hline 
$l$ & $SCC_{l}$ & $l$ & $SCC_{l}$ & $l$ & $SCC_{l}$ & $l$ & $SCC_{l}$ & $l$ & $SCC_{l}$ & $l$ & $SCC_{l}$\tabularnewline
\hline 
\hline 
1 & $\phantom{-}1.32\cdot10^{-5}$ & 11 & $-2.13\cdot10^{-7}$ & 21 & $-2.46\cdot10^{-7}$ & 31 & $-2.97\cdot10^{-7}$ & 41 & $-3.33\cdot10^{-7}$ & 51 & $\phantom{-}1.57\cdot10^{-7}$\tabularnewline
\hline 
2 & $\phantom{-}1.47\cdot10^{-7}$ & 12 & $-2.40\cdot10^{-7}$ & 22 & $-1.60\cdot10^{-7}$ & 32 & $\phantom{-}8.58\cdot10^{-8}$ & 42 & $\phantom{-}2.74\cdot10^{-7}$ & 52 & $-3.43\cdot10^{-7}$\tabularnewline
\hline 
3 & $-9.79\cdot10^{-8}$ & 13 & $-1.52\cdot10^{-7}$ & 23 & $\phantom{-}1.94\cdot10^{-7}$ & 33 & $\phantom{-}8.53\cdot10^{-8}$ & 43 & $\phantom{-}5.56\cdot10^{-10}$ & 53 & $-5.15\cdot10^{-8}$\tabularnewline
\hline 
4 & $\phantom{-}3.41\cdot10^{-8}$ & 14 & $-1.75\cdot10^{-8}$ & 24 & $\phantom{-}4.48\cdot10^{-8}$ & 34 & $\phantom{-}2.52\cdot10^{-8}$ & 44 & $\phantom{-}1.04\cdot10^{-7}$ & 54 & $\phantom{-}1.49\cdot10^{-7}$\tabularnewline
\hline 
5 & $\phantom{-}5.87\cdot10^{-9}$ & 15 & $-3.31\cdot10^{-8}$ & 25 & $\phantom{-}1.42\cdot10^{-7}$ & 35 & $-1.22\cdot10^{-7}$ & 45 & $\phantom{-}2.13\cdot10^{-8}$ & 55 & $\phantom{-}2.20\cdot10^{-7}$\tabularnewline
\hline 
6 & $-8.19\cdot10^{-8}$ & 16 & $-1.60\cdot10^{-8}$ & 26 & $-1.22\cdot10^{-7}$ & 36 & $\phantom{-}6.94\cdot10^{-8}$ & 46 & $-2.45\cdot10^{-7}$ & 56 & $\phantom{-}9.64\cdot10^{-9}$\tabularnewline
\hline 
7 & $-4.74\cdot10^{-8}$ & 17 & $\phantom{-}5.42\cdot10^{-8}$ & 27 & $-1.79\cdot10^{-7}$ & 37 & $\phantom{-}1.19\cdot10^{-7}$ & 47 & $-2.83\cdot10^{-8}$ &  & \tabularnewline
\hline 
8 & $\phantom{-}1.38\cdot10^{-7}$ & 18 & $\phantom{-}6.01\cdot10^{-8}$ & 28 & $-3.26\cdot10^{-7}$ & 38 & $-1.99\cdot10^{-8}$ & 48 & $-2.43\cdot10^{-7}$ &  & \tabularnewline
\hline 
9 & $\phantom{-}1.62\cdot10^{-7}$ & 19 & $\phantom{-}1.49\cdot10^{-7}$ & 29 & $-4.92\cdot10^{-8}$ & 39 & $\phantom{-}9.53\cdot10^{-8}$ & 49 & $-2.90\cdot10^{-7}$ &  & \tabularnewline
\hline 
10 & $\phantom{-}2.19\cdot10^{-7}$ & 20 & $-7.91\cdot10^{-8}$ & 30 & $\phantom{-}1.56\cdot10^{-8}$ & 40 & $-2.39\cdot10^{-7}$ & 50 & $\phantom{-}1.65\cdot10^{-8}$ &  & \tabularnewline
\hline 
\end{tabular}
\par\end{centering}
\caption{\label{tab:Correlations}Averaged correlations in lab~1 for the measurement
run on June 14, 2016. The statistical errors of all correlations in
this data set are $\sigma=1.55\cdot10^{-7}$ (one standard deviation).}
\end{table}

\newpage{}

\subsection{Statistical tests}

\begin{figure}
\begin{centering}
\includegraphics{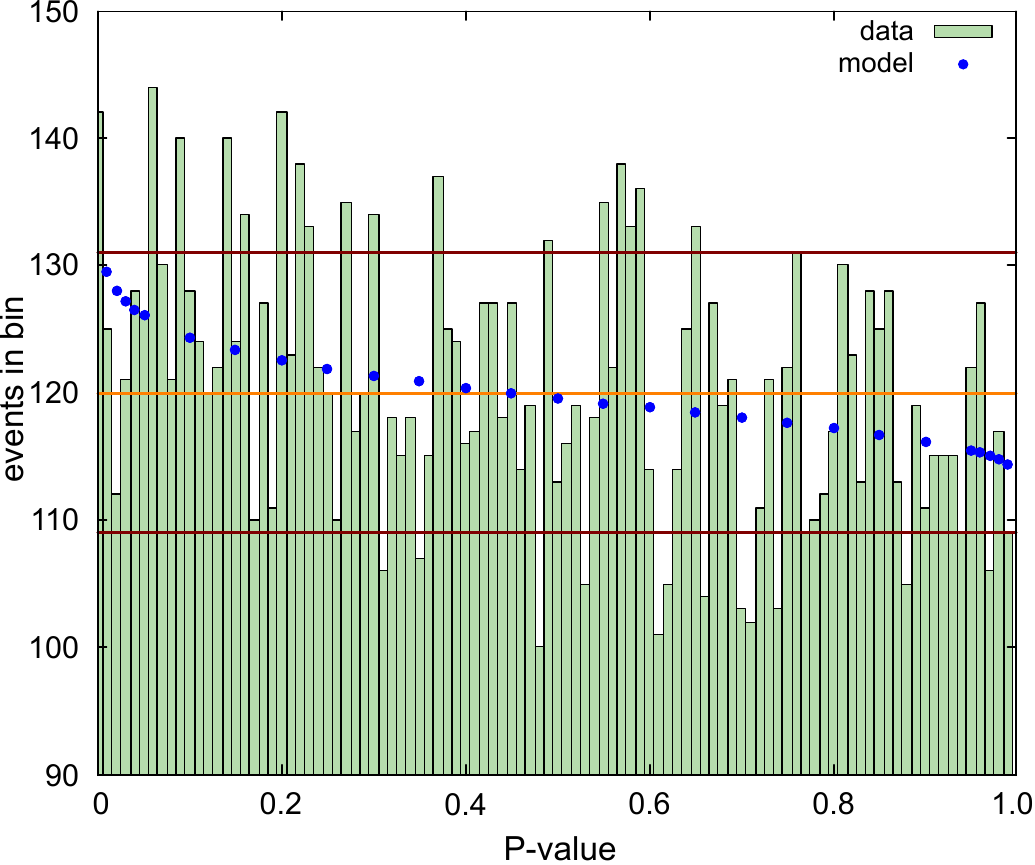}
\par\end{centering}
\caption{\label{fig:distr-P-values-model}Histogram of P-values in the serial
test smultin\_MultinomialBits with block length of $4$ bits applied
to the data from lab~1 in the measurement run on April 15, 2016.
The orange line corresponds to the average value, the red lines to
the $\pm\sigma$ interval of the bins. The blue dots result from the
model based on bias and next-neighbor correlations of this data set.}
\end{figure}

We tested the output bitsequences of the QRNGs using the statistical
test suite ``TestU01 Alphabit battery'' \cite{LEcuyer2007}. We
applied these tests to all bits collected during the measurement runs
on Apr. 15, 2016 and June 14, 2016, which are $115420$ files ($115\,\mathrm{Tb}$
in total). For each data file, all statistical tests were applied\footnote{The significance level was chosen to be $0.001$. Since we take the
uniformity of the P-value distribution as a measure of randomness,
the significance level is of no particular importance. Still, the
number of files which ``fail'' a certain test, i.e., where the P-value
is too close to $0$ or $1$ is indeed compatible with the significance
level, as expected for a uniform distribution.} and the resulting P-values for the null hypothesis of randomness
(iid bits with $Pr(q_{i}=1)=Pr(q_{i}=0)=\frac{1}{2}$) were calculated.
In the ideal case the P-values are expected to be uniformly distributed.
This was observed for all of our data and all tests in the battery
except for the test row smultin\_MultinomialBitsOver (test on uniformity
of appearance of bit chains of certain length, evaluated using overlapping
serial approach), see Table \ref{tab:Alphabit-results}. There the
P-value distribution is shifted towards $0$. To understand this behavior
we applied a related test smultin\_MultinomialBits (the same as above
but evaluated using non-overlapping serial approach) which can be
easier modeled using a noncentral $\chi^{2}$-distribution. Fig. \ref{fig:distr-P-values-model}
shows that our model which includes only the known bias and next-neighbor
correlation fits the data well. Thus the applied set of tests did
not reveal any effects in the data which are stronger than the next-neighbor
correlation. Note that these subtle effects are only visible due to
the large amount of data available. Alltogether our findings support
the thesis that the bits are in fact random and well-suited for our
application.

\begin{table}[H]
\begin{centering}
\begin{tabular}{|l||c|c|c|c|}
\hline 
 & \multicolumn{2}{c|}{April 15, 2016} & \multicolumn{2}{c|}{June 14, 2016}\tabularnewline
\hline 
Test & P-value lab 1 & P-value lab 2 & P-value lab 1 & P-value lab 2\tabularnewline
\hline 
\hline 
smultin\_MultinomialBitsOver with L = 2 & $3.83\cdot10^{-20}$ & $1.66\cdot10^{-5}$ & $6.06\cdot10^{-123}$ & $4.11\cdot10^{-44}$\tabularnewline
\hline 
smultin\_MultinomialBitsOver with L = 4 & $7.65\cdot10^{-9}$ & $1.86\cdot10^{-4}$ & $2.95\cdot10^{-58}$ & $6.92\cdot10^{-26}$\tabularnewline
\hline 
smultin\_MultinomialBitsOver with L = 8 & $0.298$ & $0.542$ & $8.14\cdot10^{-3}$ & $0.061$\tabularnewline
\hline 
smultin\_MultinomialBitsOver with L = 16 & $0.347$ & $0.086$ & $0.842$ & $0.815$\tabularnewline
\hline 
sstring\_HammingIndep with L = 16  & $0.663$ & $0.824$ & $0.065$ & $0.822$\tabularnewline
\hline 
sstring\_HammingIndep with L = 32  & $0.499$ & $0.838$ & $0.997$ & $0.043$\tabularnewline
\hline 
sstring\_HammingCorr with L = 32  & $0.275$ & $0.674$ & $0.435$ & $0.008$\tabularnewline
\hline 
swalk\_RandomWalk1 with L = 64 (Statistic H) & $0.413$ & $0.942$ & $0.045$ & $0.208$\tabularnewline
\hline 
swalk\_RandomWalk1 with L = 64 (Statistic M) & $0.166$ & $0.646$ & $0.376$ & $0.892$\tabularnewline
\hline 
swalk\_RandomWalk1 with L = 64 (Statistic J) & $0.496$ & $0.684$ & $0.506$ & $0.899$\tabularnewline
\hline 
swalk\_RandomWalk1 with L = 64 (Statistic R) & $0.092$ & $0.287$ & $0.801$ & $0.544$\tabularnewline
\hline 
swalk\_RandomWalk1 with L = 64 (Statistic C) & $0.676$ & $0.594$ & $0.712$ & $0.240$\tabularnewline
\hline 
swalk\_RandomWalk1 with L = 320 (Statistic H) & $0.963$ & $0.761$ & $0.334$ & $0.663$\tabularnewline
\hline 
swalk\_RandomWalk1 with L = 320 (Statistic M) & $0.534$ & $0.138$ & $0.575$ & $0.241$\tabularnewline
\hline 
swalk\_RandomWalk1 with L = 320 (Statistic J) & $0.487$ & $0.196$ & $0.253$ & $0.881$\tabularnewline
\hline 
swalk\_RandomWalk1 with L = 320 (Statistic R) & $0.032$ & $0.471$ & $0.291$ & $0.715$\tabularnewline
\hline 
swalk\_RandomWalk1 with L = 320 (Statistic C) & $0.941$ & $0.608$ & $0.930$ & $0.173$\tabularnewline
\hline 
\end{tabular}
\par\end{centering}
\caption{\label{tab:Alphabit-results}Results of the TestU01 Alphabit test
battery for the measurement runs on April 15, 2016 and June 14, 2016.
On each side the tests were applied to $16072$ and $41638$ files
of $1\,\mathrm{Gb}$ for the two runs, respectively. The resulting
distributions of P-values for each test were checked for uniformity
with a $\chi^{2}$-test, whose P-values are shown here.}
\end{table}

\newpage{}

\section{\label{sec:Testing-LHV-Theories}Testing LHV Theories}

The main goal of Bell experiments is to rule out all local-hidden-variable
(LHV) theories by measuring a violation of Bell's inequality. Since
real experiments can only generate a finite amount of data, even an
experiment governed by LHVs may still exhibit a violation of Bell's
inequality by chance. To account for this, one estimates the probability
that a specific violation or a more extreme one can be produced by
an experiment governed by LHVs. If this probability is small for an
experimental outcome it is fair to reject the hypothesis of LHVs with
a certain confidence. This procedure is called a \textit{null hypothesis
test} and the respective probability is called a \textit{P-value}.
For a null hypothesis test of the validity of LHV theories we have
to find the probability distribution for the $S$-values under the
assumption of LHVs. With this probability distribution we can calculate
the P-value for any measured value of $S$.

When performing such an analysis, one needs to be careful not to introduce
additional assumptions into the model to be tested. The standard way
of the evaluation of experimental data would be to assume Gaussian
distribution of measurement results. Then the P-value can be easily
calculated for the measured value of $S$ and its standard deviation.
However, this requires the assumption that the experimental tries
can be considered independent and identically distributed (\emph{iid})
which is not necessarily valid. First, the experimental parameters
may vary over the course of the experiment. Second, from the fundamental
point of view, the knowledge of the history of previous settings and
outcomes might allow for a LHV model violating Bell's inequality (\emph{memory
loophole}). Depending on the formulation of the inequality this is
indeed possible \cite{Barrett2002}, although the violation approaches
zero with increasing number of tries $N$. In any case an analysis
procedure which does not require the assumption of \emph{iid} is needed.
Ways to achieve this include modeling of the underlying stochastic
process as a \emph{martingale} \cite{Gill2003} or its formulation
as a \emph{game} \cite{Brunner2014,Elkouss2016}, as well as \emph{prediction-based-ratio}
(PBR) approach \cite{Zhang2013}.

\subsection{Defining the null hypothesis}

The first step is to formulate the null hypothesis in a way that we
can use to calculate bounds on the P-value. For this we use the CHSH
inequality as a mathematical formulation of the hypothesis and define
it for our experiment.

Our Bell experiment employs two observers in lab~1 and lab~2. On
receiving the heralding signal each side of the experiment gets an
input for the setting selection and produces a measurement outcome
(in the following we call such a process ``event''). We name the
inputs for the \textit{$i$}-th of $N$ events $a_{i}\in\left\{ \alpha,\alpha'\right\} $
for lab~1 and $b_{i}\in\left\{ \beta,\beta'\right\} $ for lab~2.
Similarly we name the measurement outcomes for this event $x_{i}$
for lab~1 and $y_{i}$ for lab~2 with $x_{i},y_{i}\in\left\{ -1,1\right\} $
(outcome $\left|\uparrow\right\rangle $ corresponding to $1$, outcome
$\left|\downarrow\right\rangle $ to $-1$). Since our experiment
employs an event-ready scheme we name the event-ready signal for this
event $h_{i}$, where $h_{i}=1$ heralds the $\Psi^{+}$-state and
$h_{i}=-1$ heralds the $\Psi^{-}$-state.

We define the functions $g^{\pm}(a,b)$ for events with the event-ready
signal $h_{i}$ for $\Psi^{+}$ or $\Psi^{-}$ that take the values
$1$ or $-1$: 
\begin{align}
g^{+}(a,b) & =-1\;\;\;\forall\:(a,b)\neq(\alpha',\beta')\\
g^{+}(a,b) & =\phantom{-}1\;\;\;\phantom{\forall}\:(a,b)=(\alpha',\beta')\nonumber \\
g^{-}(a,b) & =-1\;\;\;\forall\:(a,b)\neq(\alpha',\beta)\nonumber \\
g^{-}(a,b) & =\phantom{-}1\;\;\;\phantom{\forall}\:(a,b)=(\alpha',\beta)\nonumber 
\end{align}
With these functions we can write the CHSH inequality for each state
in the form: 
\begin{equation}
S^{\pm}=\sum_{\substack{a\in\left\{ \alpha,\alpha'\right\} \\
b\in\left\{ \beta,\beta'\right\} 
}
}g^{\pm}(a,b)\frac{^{C}N_{a,b}^{\pm}-{}^{A}N_{a,b}^{\pm}}{^{C}N_{a,b}^{\pm}+{}^{A}N_{a,b}^{\pm}}\leq2.
\end{equation}
Here $^{C}N_{a,b}^{\pm}=N_{a,b}^{\uparrow\uparrow}+N_{a,b}^{\downarrow\downarrow}$
is the number of correlated measurement outcomes ($x_{i}=y_{i}$)
and $^{A}N_{a,b}^{\pm}=N_{a,b}^{\uparrow\downarrow}+N_{a,b}^{\downarrow\uparrow}$
is the number of anticorrelated measurement outcomes ($x_{i}\neq y_{i}$),
for state $\Psi^{\pm}$ respectively. 

For large $N$ and low bias of the input random bits we can approximate
$^{C}N_{a,b}^{\pm}+{}^{A}N_{a,b}^{\pm}$ with $N^{\pm}/4$ and obtain
\begin{equation}
S^{\pm}=\frac{1}{N^{\pm}}\sum_{i=1}^{N^{\pm}}4\cdot g^{\pm}(a_{i},b_{i})x_{i}y_{i}\leq2.\label{eq:S-Mcd}
\end{equation}
This form of Bell's inequality is not susceptible to systematic violations
exploiting finite statistics shown in \cite{Barrett2002}. 

\subsection{Martingales and concentration inequalities}

In order to estimate the P-value one can employ concentration inequalities
which provide bounds on deviations of stochastic processes from their
expectation values. Here we will use the Mc Diarmid inequality \cite{McDiarmid1989}
in the following way similar to \cite{Zhang2013}. First we define
the sequences 
\begin{align}
f_{i} & =4\cdot g^{\pm}(a_{i},b_{i})x_{i}y_{i}\\
\Rightarrow & -4\leq f_{i}\leq4\nonumber 
\end{align}
and 
\begin{align}
Z_{i} & =\frac{f_{i}-2}{8}\\
\Rightarrow & -\frac{3}{4}\leq Z_{i}\leq\frac{1}{4}.\nonumber 
\end{align}
We can now write equation (\ref{eq:S-Mcd}) as 
\begin{equation}
S^{\pm}=\frac{1}{N^{\pm}}\sum_{i=1}^{N^{\pm}}f_{i}
\end{equation}
and 
\begin{equation}
\frac{1}{N^{\pm}}\sum_{i=1}^{N^{\pm}}Z_{i}=\frac{S^{\pm}-2}{8}
\end{equation}
is a measure for the violation of Bell's inequality. For any experiment
governed by LHVs saturating Bell's inequality, $Z_{i}$ is a \textit{\emph{martingale
difference sequence}} and we can use the equation (6.1) from \cite{McDiarmid1989}
\begin{equation}
Pr(\sum_{i=1}^{N^{\pm}}Z_{i}\geq tN^{\pm})\leq\left[\left[\frac{A}{A+t}\right]^{A+t}\left[\frac{\bar{A}}{\bar{A}-t}\right]^{\bar{A}-t}\right]^{N^{\pm}}\label{eq:McDiarmid}
\end{equation}
with $t=\frac{S^{\pm}-2}{8}$, $A=\frac{3}{4}$, and $\bar{A}=\frac{1}{4}$
to bound the probability of a certain violation $\delta$ or a more
extreme one under assumption of LHVs, thereby obtaining the P-value
$P_{m}$:
\begin{equation}
P_{m}=Pr(S-2\geq\delta)\leq\left[\frac{1}{\left(1+\frac{\delta}{6}\right)^{\frac{3}{4}+\frac{\delta}{8}}\cdot\left(1-\frac{\delta}{2}\right)^{\frac{1}{4}-\frac{\delta}{8}}}\right]^{N}.\label{eq:McDiarmid-explicit}
\end{equation}
 This bound is also valid for all LHV models which do not saturate
Bell's inequality, in which case the process is a supermartingale.

The limit of the $S$-value achievable with LHVs increases if the
random bits are not ideal (partially predictable). The deviations
from perfect unpredictability $\tau_{a},\tau_{b}\in\left[-\frac{1}{2},\frac{1}{2}\right]$
on the two sides are defined $\forall i$ by 
\begin{align}
\frac{1}{2}-\tau_{a}\text{\ensuremath{\leq}}Pr(a_{i}=\alpha|H_{i}) & \leq\frac{1}{2}+\tau_{a}\nonumber \\
\frac{1}{2}-\tau_{b}\text{\ensuremath{\leq}}Pr(b_{i}=\beta|H_{i}) & \leq\frac{1}{2}+\tau_{b}\label{eq:Predictability}
\end{align}
where $H_{i}$ is the common history of the experiment at the $i$-th
event, i.e. the complete information available to the two observers
which can be used for predicting the setting choice on the other side.

For simplicity we set $\tau=\max(\left|\tau_{a}\right|,\left|\tau_{b}\right|)$.
We note that the predictability may depend on the history of the experiment
$H$

To calculate the effect of the predictability on the $S$-parameter
we consider the following. For example let us assume that for a certain
attempt $i$ where the atomic state $\left|\Psi^{+}\right\rangle $
was prepared, the probabilities of the setting inputs are $Pr(a_{i}=\alpha)=\frac{1}{2}+\tau$
and $Pr(b_{i}=\beta)=\frac{1}{2}+\tau$. This means that in this situation
the probability of $(\alpha,\beta)$ input is the highest and of the
$(\alpha',\beta')$ input is the lowest. An LHV strategy (of the $16$
possible) which would maximize the expectation value of $S$ for this
attempt would be to produce anticorrelations for all input combinations.
Then one obtains for this expectation value
\begin{multline*}
\sum_{\substack{a\in\left\{ \alpha,\alpha'\right\} \\
b\in\left\{ \beta,\beta'\right\} 
}
}4\cdot g^{+}(a,b)\cdot x_{i}y_{i}\cdot Pr(a_{i}=a)Pr(b_{i}=b)=\\
=4\Big((-1)(-1)(\frac{1}{2}+\tau)^{2}+(-1)(-1)(\frac{1}{2}+\tau)(\frac{1}{2}-\tau)\\
+(-1)(-1)(\frac{1}{2}-\tau)(\frac{1}{2}+\tau)+(+1)(-1)(\frac{1}{2}-\tau)^{2}\Big)=\\
=4\left(2\tau+2(\frac{1}{4}-\tau^{2})\right)=2+8(\tau-\tau^{2}).
\end{multline*}
As is easy to verify, for any combination of input setting probabilities
and each atomic state there exists an LHV strategy achieving this
value. Thus, by optimally switching strategies event for event, one
obtains the expectation value 
\[
S\leq2+8(\tau-\tau^{2}).
\]
We consequentially use $t=\frac{S^{\pm}-2}{8}-(\tau-\tau^{2})$ ,
$A=\frac{3}{4}+(\tau-\tau^{2})$, and $\bar{A}=\frac{1}{4}-(\tau-\tau^{2})$
in (\ref{eq:McDiarmid}) for calculation of P-values.

To calculate the P-value for the combined data of $\Psi^{+}$- and
$\Psi^{-}$- states we use

\begin{equation}
S=\frac{1}{N}\sum_{i=1}^{N}4\cdot\left[\frac{1+h_{i}}{2}g^{+}(a_{i},b_{i})x_{i}y_{i}+\frac{1-h_{i}}{2}g^{-}(a_{i},b_{i})x_{i}y_{i}\right]\leq2.\label{eq:S-combined-McD}
\end{equation}

\subsection{Game formalism}

An alternative approach is to formulate a CHSH experiment as a game
\cite{Elkouss2016}. Here, LHV are represented by two parties which
have to generate correlations or anticorrelations based on random
inputs, where only the local input is known to each party during a
game round. Otherwise they may employ any strategy which also may
be adapted during the course of the game and are allowed to communicate
between the rounds. To win the game they need to produce the right
correlations (three anticorrelations, one correlation, depending on
the input) maximal number of times.

To describe this formally, we define the functions $w_{i}^{+}$ for
$\Psi^{+}$-events and $w_{i}^{-}$ for $\Psi^{-}$-events: 
\begin{equation}
w_{i}^{\pm}=\frac{\left|g^{\pm}(a_{i},b_{i})+x_{i}y_{i}\right|}{2}.
\end{equation}
If $w_{i}^{\pm}=1$ the game is won for this round. Thus the total
number of rounds won for each state is $W^{\pm}=\sum_{i=1}^{N}\frac{1\pm h_{i}}{2}w_{i}^{\pm}$.
For LHV theories the probability of winning a single round of a CHSH-game
is \cite{Elkouss2016} 
\begin{equation}
Pr(w_{i}^{\pm}=1)\leq\xi=3/4+(\tau-\tau^{2}).
\end{equation}
Note that, if $\tau$ depends on the history of the experiment (Eq.
(\ref{eq:Predictability})), the winning probability will also depend
on this history. Now we can calculate the probability $Pr(W,N)$ of
winning at least $W$ times in $N$ rounds: 
\begin{equation}
P_{g}\leq Pr(W,N)=\sum_{j=W}^{N}\binom{N}{j}\xi^{j}(1-\xi)^{N-j}.\label{eq:P-value-game}
\end{equation}
With the number of wins $W^{\pm}$ and number of events $N^{\pm}$
for each atomic state we can calculate a P-value $P_{g}$ for each
state individually. To calculate a combined P-value for the complete
experiment we define the function for a win for $\Psi^{+}$ and $\Psi^{-}$
events 
\begin{equation}
w_{i}=\frac{1+h_{i}}{2}\frac{\left|g^{+}(a_{i},b_{i})+x_{i}y_{i}\right|}{2}+\frac{1-h_{i}}{2}\frac{\left|g^{-}(a_{i},b_{i})+x_{i}y_{i}\right|}{2}.
\end{equation}
The total number of wins for $\Psi^{+}$ and $\Psi^{-}$ is $W=W^{+}+W^{-}$
and can be put in Eq. (\ref{eq:P-value-game}). 

\newpage{}

\section{Avoiding Expectation Bias}

An important property of any scientific experiment should be impartiality.
This implies that the assessment of the results has to be based on
objective criteria only, devoid of any expectation on the outcome.
If no special care is taken about this, results can easily become
biased towards the expectation (see, e.g., \cite{Jeng2006}). This
can happen, e.g., by conscious or unconscious discarding of data which
apparently do not fit the expected value. Publishing predominantly
positive results can lead to a distorted picture in the literature,
known as the ``publication bias''. Vice versa, distorted values
in the literature may influence new experiments.

The number of parameters in a complex experiment can be large making
it difficult to define a complete set of objective criteria for a
decision whether a certain experimental run is valid or not. However,
one must not decide on its validity by looking at the result. This
also prohibits discarding parts of the data where the result deviates
from the expectation or stopping the run prematurely when the result
appears acceptable. Doing so will lead to apparently ``better''
results (``P-value hacking'').

To account for this problem we defined a list of rules before a run
is started (we admit that completely avoiding bias is extremely difficult
and our measures might not be complete):
\begin{itemize}
\item The number of events to be accumulated is fixed beforehand.
\item The acquisition procedure is fixed beforehand. This includes all acceptance
time-windows (two-photon coincidence, CEM detections).
\item The analysis procedure is fixed beforehand. This includes the calculation
of $S$ and P-values.
\item Exclusion of events during the experimental run is based only on the
two following criteria:

\begin{itemize}
\item laser stability: on each side all stabilized lasers are fed into a
scanning Fabry-Perot resonator whose output is measured with a photodiode.
The resulting spectrum is represented with an oscilloscope and monitored
by a camera. This allows us to (manually) determine the time when
a malfunction appeared and to exclude all events between this time
and until the problem is fixed. Malfunctions of most lasers will yield
no events (as no atoms will be loaded), however problems with the
readout laser reduce the fidelity and such events have to be excluded.
\item CEMs: a high-voltage breakthrough in the detector system can lead
to a shutdown. In this case all events are automatically excluded
until the problem is fixed.
\end{itemize}
\item The maintenance procedure is performed every 24 hours and is limited
to:

\begin{itemize}
\item check of the laser system (frequency stabilization and optical power
at all relevant positions),
\item compensation of magnetic fields (for all $3$ axes with a precision
of $0.5\,\mathrm{mG}$),
\item minimization of polarization rotation in the fibers of the beam splitter
in the BSM arrangement including the $5\,\mathrm{m}$ fiber connecting
trap~1 to the BSM. Together with the automatic polarization compensation
procedure of the $700\,\mathrm{m}$ fiber from trap~2 this ensures
that there is no polarization rotation between different inputs and
outputs of the BS in the two-photon interference process.
\end{itemize}
\end{itemize}
\newpage{}

\section{Public Measurement Run}

On top of obtaining a conclusive violation of Bell's inequality, an
additional goal of our project was to perform an open scientific experiment.
This includes defining the rules in advance to avoid expectation bias,
as well as making all data available to the public during the whole
course of the experiment. For this purpose we have set up a web server
\url{http://bellexp.quantum.physik.uni-muenchen.de}. All incoming
data and other relevant information is presented there in real-time.
Important information concerning the measurement is logged and distributed
via the Twitter account \emph{@munichbellexp}.

The public run was started on June 14, 2016 with the goal to collect
$5000$ events for each of the two prepared atomic states. The run
took $10$ days including daily maintenance stops. While the obtained
violation (Tab. \ref{tab:data-20160614}) is weaker than in other
runs, it is still significant. Note that the situation where a certain
measurement run yields results below average is to be expected, the
same way results that appear above the average can be obtained due
to purely statistical effects.\newpage{}

\section{Data Tables}

Experimental data is available online at

\url{http://bellexp.quantum.physik.uni-muenchen.de}.

\subsection{Run on Nov. 27, 2015}

\begin{table}[h]
\begin{centering}
{\footnotesize{}}%
\begin{tabular}{|c||c|c|c|c||c||c|}
\hline 
\multicolumn{7}{|c|}{{\footnotesize{}Nov. 27, 2015, $\Psi^{+}$}}\tabularnewline
\hline 
\hline 
{\footnotesize{}$\alpha,\beta$} & {\footnotesize{}$\uparrow\uparrow$} & {\footnotesize{}$\uparrow\downarrow$} & {\footnotesize{}$\downarrow\uparrow$} & {\footnotesize{}$\downarrow\downarrow$} & {\footnotesize{}total } & {\footnotesize{}$\left\langle \sigma_{\alpha}\sigma_{\beta}\right\rangle $}\tabularnewline
\hline 
\hline 
{\footnotesize{}0,-45 } & {\footnotesize{}4} & {\footnotesize{}16} & {\footnotesize{}21} & {\footnotesize{}4} & {\footnotesize{}45} & {\footnotesize{}$-0.644\pm0.114$ }\tabularnewline
\hline 
{\footnotesize{}0,\hphantom{-}45 } & {\footnotesize{}4} & {\footnotesize{}12} & {\footnotesize{}13} & {\footnotesize{}4} & {\footnotesize{}33} & {\footnotesize{}$-0.515\pm0.149$}\tabularnewline
\hline 
{\footnotesize{}90,-45 } & {\footnotesize{}3} & {\footnotesize{}24} & {\footnotesize{}11} & {\footnotesize{}2} & {\footnotesize{}40} & {\footnotesize{}$-0.750\pm0.105$}\tabularnewline
\hline 
{\footnotesize{}90,\hphantom{-}45 } & {\footnotesize{}10} & {\footnotesize{}4 } & {\footnotesize{}8} & {\footnotesize{}10} & {\footnotesize{}32} & {\footnotesize{}$\phantom{-}0.250\pm0.171$}\tabularnewline
\hline 
\hline 
{\footnotesize{}$S$} & \multicolumn{4}{c||}{} & {\footnotesize{}150} & {\footnotesize{}$\phantom{-}\mathbf{2.160\pm0.279}$}\tabularnewline
\hline 
\hline 
{\footnotesize{}$P_{m}$} & \multicolumn{5}{c||}{} & {\footnotesize{}$0.7009$}\tabularnewline
\hline 
{\footnotesize{}$P_{g}$} & \multicolumn{5}{c||}{{\footnotesize{} 117 wins }} & {\footnotesize{}$0.2328$}\tabularnewline
\hline 
\end{tabular}{\footnotesize{}\hfill{}}%
\begin{tabular}{|c||c|c|c|c||c||c|}
\hline 
\multicolumn{7}{|c|}{{\footnotesize{}Nov. 27, 2015, $\Psi^{-}$}}\tabularnewline
\hline 
\hline 
{\footnotesize{}$\alpha,\beta$} & {\footnotesize{}$\uparrow\uparrow$} & {\footnotesize{}$\uparrow\downarrow$} & {\footnotesize{}$\downarrow\uparrow$} & {\footnotesize{}$\downarrow\downarrow$} & {\footnotesize{}total } & {\footnotesize{}$\left\langle \sigma_{\alpha}\sigma_{\beta}\right\rangle $}\tabularnewline
\hline 
\hline 
{\footnotesize{}0,-45 } & {\footnotesize{}4} & {\footnotesize{}11} & {\footnotesize{}17} & {\footnotesize{}2} & {\footnotesize{}34} & {\footnotesize{}$-0.647\pm0.131$ }\tabularnewline
\hline 
{\footnotesize{}0,\hphantom{-}45 } & {\footnotesize{}4} & {\footnotesize{}16} & {\footnotesize{}13} & {\footnotesize{}3} & {\footnotesize{}36} & {\footnotesize{}$-0.611\pm0.132$}\tabularnewline
\hline 
{\footnotesize{}90,-45 } & {\footnotesize{}22} & {\footnotesize{}4} & {\footnotesize{}2} & {\footnotesize{}10} & {\footnotesize{}38} & {\footnotesize{}$\phantom{-}0.684\pm0.118$}\tabularnewline
\hline 
{\footnotesize{}90,\hphantom{-}45 } & {\footnotesize{}4} & {\footnotesize{}19} & {\footnotesize{}16} & {\footnotesize{}3} & {\footnotesize{}42} & {\footnotesize{}$-0.666\pm0.115$}\tabularnewline
\hline 
\hline 
{\footnotesize{}$S$} & \multicolumn{4}{c||}{} & {\footnotesize{}150} & {\footnotesize{}$\phantom{-}\mathbf{2.609\pm0.252}$}\tabularnewline
\hline 
\hline 
{\footnotesize{}$P_{m}$} & \multicolumn{5}{c||}{} & {\footnotesize{}$0.0814$}\tabularnewline
\hline 
{\footnotesize{}$P_{g}$} & \multicolumn{5}{c||}{{\footnotesize{}124 wins }} & {\footnotesize{}$0.0170$}\tabularnewline
\hline 
\end{tabular}
\par\end{centering}{\footnotesize \par}
\caption{Experimental data from the run on Nov. 27, 2015.}
\end{table}

\begin{table}[h]
\begin{centering}
\begin{tabular}{|c|c|}
\hline 
method & S-value\tabularnewline
\hline 
\hline 
weighted arithmetic mean & $2.407\pm0.184$\tabularnewline
\hline 
event based & $2.415\pm0.185$\tabularnewline
\hline 
\end{tabular}
\par\end{centering}
\vspace{0.5cm}

\begin{centering}
\begin{tabular}{|c|c|}
\hline 
method & P-value\tabularnewline
\hline 
\hline 
$P_{m}$ with combined S-value & $0.0958$\tabularnewline
\hline 
$P_{g}$ with $241$ wins of $300$ & $0.0186$\tabularnewline
\hline 
\end{tabular}
\par\end{centering}
\caption{Evaluation of combined $S$- and P-values for the run of Nov. 27,
2015.}
\end{table}

\newpage{}

\subsection{Run on April 7, 2016}

\begin{table}[h]
\begin{centering}
{\footnotesize{}}%
\begin{tabular}{|c||c|c|c|c||c||c|}
\hline 
\multicolumn{7}{|c|}{{\footnotesize{}April 7, 2016, $\Psi^{+}$}}\tabularnewline
\hline 
\hline 
{\footnotesize{}$\alpha,\beta$} & {\footnotesize{}$\uparrow\uparrow$} & {\footnotesize{}$\uparrow\downarrow$} & {\footnotesize{}$\downarrow\uparrow$} & {\footnotesize{}$\downarrow\downarrow$} & {\footnotesize{}total } & {\footnotesize{}$\left\langle \sigma_{\alpha}\sigma_{\beta}\right\rangle $}\tabularnewline
\hline 
\hline 
{\footnotesize{}0,-45 } & {\footnotesize{}60} & {\footnotesize{}182} & {\footnotesize{}164} & {\footnotesize{}47} & {\footnotesize{}453 } & {\footnotesize{}$-0.528\pm0.040$ }\tabularnewline
\hline 
{\footnotesize{}0,\hphantom{-}45 } & {\footnotesize{}50} & {\footnotesize{}182} & {\footnotesize{}196} & {\footnotesize{}47} & {\footnotesize{}475 } & {\footnotesize{}$-0.592\pm0.037$}\tabularnewline
\hline 
{\footnotesize{}90,-45 } & {\footnotesize{}51} & {\footnotesize{}168} & {\footnotesize{}182} & {\footnotesize{}29} & {\footnotesize{}430 } & {\footnotesize{}$-0.628\pm0.038$}\tabularnewline
\hline 
{\footnotesize{}90,\hphantom{-}45 } & {\footnotesize{}195} & {\footnotesize{}60} & {\footnotesize{}62} & {\footnotesize{}159} & {\footnotesize{}476 } & {\footnotesize{}$\phantom{-}0.487\pm0.040$}\tabularnewline
\hline 
\hline 
{\footnotesize{}$S$} & \multicolumn{4}{c||}{} & {\footnotesize{}1834 } & {\footnotesize{}$\mathbf{\phantom{-}2.234\pm0.077}$}\tabularnewline
\hline 
\hline 
{\footnotesize{}$P_{m}$} & \multicolumn{5}{c||}{} & {\footnotesize{}$0.0194$}\tabularnewline
\hline 
{\footnotesize{}$P_{g}$} & \multicolumn{5}{c||}{{\footnotesize{}1428 wins}} & {\footnotesize{}$2.755\cdot10^{-03}$}\tabularnewline
\hline 
\end{tabular}{\footnotesize{}\hfill{}}%
\begin{tabular}{|c||c|c|c|c||c||c|}
\hline 
\multicolumn{7}{|c|}{{\footnotesize{}April 7, 2016, $\Psi^{-}$}}\tabularnewline
\hline 
\hline 
{\footnotesize{}$\alpha,\beta$} & {\footnotesize{}$\uparrow\uparrow$} & {\footnotesize{}$\uparrow\downarrow$} & {\footnotesize{}$\downarrow\uparrow$} & {\footnotesize{}$\downarrow\downarrow$} & {\footnotesize{}total } & {\footnotesize{}$\left\langle \sigma_{\alpha}\sigma_{\beta}\right\rangle $}\tabularnewline
\hline 
\hline 
{\footnotesize{}0,-45 } & {\footnotesize{}62} & {\footnotesize{}192} & {\footnotesize{}175} & {\footnotesize{}73} & {\footnotesize{}502 } & {\footnotesize{}$-0.462\pm0.040$ }\tabularnewline
\hline 
{\footnotesize{}0,\hphantom{-}45 } & {\footnotesize{}31} & {\footnotesize{}189} & {\footnotesize{}184} & {\footnotesize{}29} & {\footnotesize{}433 } & {\footnotesize{}$-0.723\pm0.033$}\tabularnewline
\hline 
{\footnotesize{}90,-45 } & {\footnotesize{}213} & {\footnotesize{}38} & {\footnotesize{}44} & {\footnotesize{}187} & {\footnotesize{}482 } & {\footnotesize{}$\phantom{-}0.660\pm0.034$}\tabularnewline
\hline 
{\footnotesize{}90,\hphantom{-}45 } & {\footnotesize{}77} & {\footnotesize{}166} & {\footnotesize{}165} & {\footnotesize{}52} & {\footnotesize{}460 } & {\footnotesize{}$-0.439\pm0.042$}\tabularnewline
\hline 
\hline 
{\footnotesize{}$S$} & \multicolumn{4}{c||}{} & {\footnotesize{}1877 } & {\footnotesize{}$\mathbf{\phantom{-}2.284\pm0.075}$}\tabularnewline
\hline 
\hline 
{\footnotesize{}$P_{m}$} & \multicolumn{5}{c||}{} & {\footnotesize{}$3.490\cdot10^{-03}$}\tabularnewline
\hline 
{\footnotesize{}$P_{g}$} & \multicolumn{5}{c||}{{\footnotesize{}1471 wins}} & {\footnotesize{}$4.317\cdot10^{-04}$}\tabularnewline
\hline 
\end{tabular}
\par\end{centering}{\footnotesize \par}
\caption{Experimental data from the run on April 7, 2016.}
\end{table}

\begin{center}
\begin{table}[h]
\begin{centering}
\begin{tabular}{|c|c|}
\hline 
method & S-value\tabularnewline
\hline 
\hline 
weighted arithmetic mean & $2.260\pm0.0537$\tabularnewline
\hline 
event based & $2.256\pm0.0537$\tabularnewline
\hline 
\end{tabular}
\par\end{centering}
\medskip{}

\begin{centering}
\begin{tabular}{|c|c|}
\hline 
method & P-value\tabularnewline
\hline 
\hline 
$P_{m}$ with combined S-value & $7.261\cdot10^{-5}$\tabularnewline
\hline 
$P_{g}$ with $2899$ wins of $3711$ & $7.027\cdot10^{-6}$\tabularnewline
\hline 
\end{tabular}
\par\end{centering}
\caption{Evaluation of combined $S$- and P-values for the run of April 7,
2016.}
\end{table}
\par\end{center}

\newpage{}

\subsection{Run on Apr. 15, 2016}

\begin{table}[h]
\begin{centering}
{\footnotesize{}}%
\begin{tabular}{|c||c|c|c|c||c||c|}
\hline 
\multicolumn{7}{|c|}{{\footnotesize{}Apr. 15, 2016, $\Psi^{+}$}}\tabularnewline
\hline 
\hline 
{\footnotesize{}$\alpha,\beta$} & {\footnotesize{}$\uparrow\uparrow$} & {\footnotesize{}$\uparrow\downarrow$} & {\footnotesize{}$\downarrow\uparrow$} & {\footnotesize{}$\downarrow\downarrow$} & {\footnotesize{}total } & {\footnotesize{}$\left\langle \sigma_{\alpha}\sigma_{\beta}\right\rangle $}\tabularnewline
\hline 
\hline 
{\footnotesize{}0,-45 } & {\footnotesize{}154 } & {\footnotesize{}483 } & {\footnotesize{}471 } & {\footnotesize{}135 } & {\footnotesize{}1243 } & {\footnotesize{}$-0.535\pm0.024$ }\tabularnewline
\hline 
{\footnotesize{}0,\hphantom{-}45 } & {\footnotesize{}135 } & {\footnotesize{}471 } & {\footnotesize{}507 } & {\footnotesize{}107 } & {\footnotesize{}1220 } & {\footnotesize{}$-0.603\pm0.023$}\tabularnewline
\hline 
{\footnotesize{}90,-45 } & {\footnotesize{}134 } & {\footnotesize{}499 } & {\footnotesize{}513 } & {\footnotesize{}117 } & {\footnotesize{}1263 } & {\footnotesize{}$-0.603\pm0.022$}\tabularnewline
\hline 
{\footnotesize{}90,\hphantom{-}45 } & {\footnotesize{}489} & {\footnotesize{}160 } & {\footnotesize{}182 } & {\footnotesize{}443 } & {\footnotesize{}1274 } & {\footnotesize{}$\phantom{-}0.463\pm0.025$}\tabularnewline
\hline 
\hline 
{\footnotesize{}$S$} & \multicolumn{4}{c||}{} & {\footnotesize{}5000 } & {\footnotesize{}$\mathbf{\phantom{-}2.204\pm0.047}$}\tabularnewline
\hline 
\hline 
{\footnotesize{}$P_{m}$} & \multicolumn{5}{c||}{} & {\footnotesize{}$2.611\cdot10^{-04}$}\tabularnewline
\hline 
{\footnotesize{}$P_{g}$} & \multicolumn{5}{c||}{{\footnotesize{}3876 wins}} & {\footnotesize{}$2.643\cdot10^{-05}$}\tabularnewline
\hline 
\end{tabular}{\footnotesize{}\hfill{}}%
\begin{tabular}{|c||c|c|c|c||c||c|}
\hline 
\multicolumn{7}{|c|}{{\footnotesize{}Apr. 15, 2016, $\Psi^{-}$}}\tabularnewline
\hline 
\hline 
{\footnotesize{}$\alpha,\beta$} & {\footnotesize{}$\uparrow\uparrow$} & {\footnotesize{}$\uparrow\downarrow$} & {\footnotesize{}$\downarrow\uparrow$} & {\footnotesize{}$\downarrow\downarrow$} & {\footnotesize{}total } & {\footnotesize{}$\left\langle \sigma_{\alpha}\sigma_{\beta}\right\rangle $}\tabularnewline
\hline 
\hline 
{\footnotesize{}0,-45 } & {\footnotesize{}168 } & {\footnotesize{}443} & {\footnotesize{}536 } & {\footnotesize{}149 } & {\footnotesize{}1296 } & {\footnotesize{}$-0.511\pm0.024$ }\tabularnewline
\hline 
{\footnotesize{}0,\hphantom{-}45 } & {\footnotesize{}122 } & {\footnotesize{}492} & {\footnotesize{}510 } & {\footnotesize{}117 } & {\footnotesize{}1241 } & {\footnotesize{}$-0.615\pm0.022$}\tabularnewline
\hline 
{\footnotesize{}90,-45 } & {\footnotesize{}535} & {\footnotesize{}115 } & {\footnotesize{}128 } & {\footnotesize{}461 } & {\footnotesize{}1239 } & {\footnotesize{}$\phantom{-}0.608\pm0.023$}\tabularnewline
\hline 
{\footnotesize{}90,\hphantom{-}45 } & {\footnotesize{}172 } & {\footnotesize{}439 } & {\footnotesize{}483 } & {\footnotesize{}130 } & {\footnotesize{}1224 } & {\footnotesize{}$-0.507\pm0.025$}\tabularnewline
\hline 
\hline 
{\footnotesize{}$S$} & \multicolumn{4}{c||}{} & {\footnotesize{}5000 } & {\footnotesize{}$\mathbf{\phantom{-}2.240\pm0.047}$}\tabularnewline
\hline 
\hline 
{\footnotesize{}$P_{m}$} & \multicolumn{5}{c||}{} & {\footnotesize{}$8.4437\cdot10^{-06}$}\tabularnewline
\hline 
{\footnotesize{}$P_{g}$} & \multicolumn{5}{c||}{{\footnotesize{}3899 wins}} & {\footnotesize{}$7.397\cdot10^{-07}$}\tabularnewline
\hline 
\end{tabular}
\par\end{centering}{\footnotesize \par}
\caption{Experimental data from the run on Apr. 15, 2016.}
\end{table}

\begin{center}
\begin{table}[h]
\begin{centering}
\begin{tabular}{|c|c|}
\hline 
method & S-value\tabularnewline
\hline 
\hline 
weighted arithmetic mean & $2.222\pm0.0332$\tabularnewline
\hline 
event based & $2.221\pm0.0332$\tabularnewline
\hline 
\end{tabular}
\par\end{centering}
\medskip{}

\begin{centering}
\begin{tabular}{|c|c|}
\hline 
method & P-value\tabularnewline
\hline 
\hline 
$P_{m}$ with combined S-value & $2.569\cdot10^{-9}$\tabularnewline
\hline 
$P_{g}$ with $7775$ wins of $10000$ & $1.739\cdot10^{-10}$\tabularnewline
\hline 
\end{tabular}
\par\end{centering}
\caption{Evaluation of combined $S$- and P-values for the run of Apr. 15,
2016.}
\end{table}
\newpage{}
\par\end{center}

\subsection{Run on June 14, 2016}

\begin{table}[h]
\begin{centering}
{\footnotesize{}}%
\begin{tabular}{|c||c|c|c|c||c||c|}
\hline 
\multicolumn{7}{|c|}{{\footnotesize{}June 14, 2016, $\Psi^{+}$}}\tabularnewline
\hline 
\hline 
{\footnotesize{}$\alpha,\beta$} & {\footnotesize{}$\uparrow\uparrow$} & {\footnotesize{}$\uparrow\downarrow$} & {\footnotesize{}$\downarrow\uparrow$} & {\footnotesize{}$\downarrow\downarrow$} & {\footnotesize{}total } & {\footnotesize{}$\left\langle \sigma_{\alpha}\sigma_{\beta}\right\rangle $}\tabularnewline
\hline 
\hline 
{\footnotesize{}0,-45 } & {\footnotesize{}118} & {\footnotesize{}483} & {\footnotesize{}510} & {\footnotesize{}146} & {\footnotesize{}1257 } & {\footnotesize{}$-0.580\pm0.023$ }\tabularnewline
\hline 
{\footnotesize{}0,\hphantom{-}45 } & {\footnotesize{}144} & {\footnotesize{}482} & {\footnotesize{}450} & {\footnotesize{}185} & {\footnotesize{}1261 } & {\footnotesize{}$-0.478\pm0.025$}\tabularnewline
\hline 
{\footnotesize{}90,-45 } & {\footnotesize{}161} & {\footnotesize{}441} & {\footnotesize{}427} & {\footnotesize{}173} & {\footnotesize{}1202 } & {\footnotesize{}$-0.444\pm0.026$}\tabularnewline
\hline 
{\footnotesize{}90,\hphantom{-}45 } & {\footnotesize{}506} & {\footnotesize{}158} & {\footnotesize{}127} & {\footnotesize{}489} & {\footnotesize{}1280 } & {\footnotesize{}$\phantom{-}0.555\pm0.023$}\tabularnewline
\hline 
\hline 
{\footnotesize{}$S$} & \multicolumn{4}{c||}{} & {\footnotesize{}5000 } & {\footnotesize{}$\mathbf{\phantom{-}2.057\pm0.048}$}\tabularnewline
\hline 
\hline 
{\footnotesize{}$P_{m}$} & \multicolumn{5}{c||}{} & {\footnotesize{}$0.5205$}\tabularnewline
\hline 
{\footnotesize{}$P_{g}$} & \multicolumn{5}{c||}{{\footnotesize{}3788 wins}} & {\footnotesize{}$0.1306$}\tabularnewline
\hline 
\end{tabular}{\footnotesize{}\hfill{}}%
\begin{tabular}{|c||c|c|c|c||c||c|}
\hline 
\multicolumn{7}{|c|}{{\footnotesize{}June 14, 2016, $\Psi^{-}$}}\tabularnewline
\hline 
\hline 
{\footnotesize{}$\alpha,\beta$} & {\footnotesize{}$\uparrow\uparrow$} & {\footnotesize{}$\uparrow\downarrow$} & {\footnotesize{}$\downarrow\uparrow$} & {\footnotesize{}$\downarrow\downarrow$} & {\footnotesize{}total } & {\footnotesize{}$\left\langle \sigma_{\alpha}\sigma_{\beta}\right\rangle $}\tabularnewline
\hline 
\hline 
{\footnotesize{}0,-45 } & {\footnotesize{}133} & {\footnotesize{}533} & {\footnotesize{}537} & {\footnotesize{}105} & {\footnotesize{}1308 } & {\footnotesize{}$-0.636\pm0.021$ }\tabularnewline
\hline 
{\footnotesize{}0,\hphantom{-}45 } & {\footnotesize{}162} & {\footnotesize{}466} & {\footnotesize{}410} & {\footnotesize{}207} & {\footnotesize{}1245 } & {\footnotesize{}$-0.407\pm0.026$}\tabularnewline
\hline 
{\footnotesize{}90,-45 } & {\footnotesize{}431} & {\footnotesize{}159 } & {\footnotesize{}160 } & {\footnotesize{}454} & {\footnotesize{}1204 } & {\footnotesize{}$\phantom{-}0.470\pm0.025$}\tabularnewline
\hline 
{\footnotesize{}90,\hphantom{-}45 } & {\footnotesize{}104} & {\footnotesize{}523} & {\footnotesize{}484} & {\footnotesize{}132} & {\footnotesize{}1243 } & {\footnotesize{}$-0.620\pm0.022$}\tabularnewline
\hline 
\hline 
{\footnotesize{}$S$} & \multicolumn{4}{c||}{} & {\footnotesize{}5000 } & {\footnotesize{}$\mathbf{2.134\pm0.048}$}\tabularnewline
\hline 
\hline 
{\footnotesize{}$P_{m}$} & \multicolumn{5}{c||}{} & {\footnotesize{}$0.0201$}\tabularnewline
\hline 
{\footnotesize{}$P_{g}$} & \multicolumn{5}{c||}{{\footnotesize{}3838 wins}} & {\footnotesize{}$2.752\cdot10^{-3}$}\tabularnewline
\hline 
\end{tabular}
\par\end{centering}{\footnotesize \par}
\caption{\label{tab:data-20160614}Experimental data from the public run on
June 14, 2016.}
\end{table}

\begin{center}
\begin{table}[h]
\begin{centering}
\begin{tabular}{|c|c|}
\hline 
method & S-value\tabularnewline
\hline 
\hline 
weighted arithmetic mean & $2.096\pm0.0340$\tabularnewline
\hline 
event based & $2.096\pm0.0340$\tabularnewline
\hline 
\end{tabular}
\par\end{centering}
\medskip{}

\begin{centering}
\begin{tabular}{|c|c|}
\hline 
method & P-value\tabularnewline
\hline 
\hline 
$P_{m}$ with combined S-value & $0.0287$\tabularnewline
\hline 
$P_{g}$ with $7626$ wins of $10000$ & $2.818\cdot10^{-3}$\tabularnewline
\hline 
\end{tabular}
\par\end{centering}
\caption{Evaluation of combined $S$- and P-values for the run of June 14,
2016.}
\end{table}
\newpage{}
\par\end{center}

\subsection{Combined data of all runs}

For completeness we also provide the data of all runs between Nov.
27, 2015 and June 24, 2016. These also include test and calibration
runs, no data was sorted out.

\begin{table}[h]
\begin{centering}
{\footnotesize{}}%
\begin{tabular}{|c||c|c|c|c||c||c|}
\hline 
\multicolumn{7}{|c|}{{\footnotesize{}Nov. 11, 2015 - June 24, 2016, $\Psi^{+}$}}\tabularnewline
\hline 
\hline 
{\footnotesize{}$\alpha,\beta$} & {\footnotesize{}$\uparrow\uparrow$} & {\footnotesize{}$\uparrow\downarrow$} & {\footnotesize{}$\downarrow\uparrow$} & {\footnotesize{}$\downarrow\downarrow$} & {\footnotesize{}total } & {\footnotesize{}$\left\langle \sigma_{\alpha}\sigma_{\beta}\right\rangle $}\tabularnewline
\hline 
\hline 
{\footnotesize{}0,-45 } & {\footnotesize{}778} & {\footnotesize{}2621} & {\footnotesize{}2770} & {\footnotesize{}804} & {\footnotesize{}6973} & {\footnotesize{}$-0.546\pm0.010$}\tabularnewline
\hline 
{\footnotesize{}0,\hphantom{-}45 } & {\footnotesize{}809} & {\footnotesize{}2629} & {\footnotesize{}2708} & {\footnotesize{}816} & {\footnotesize{}6962} & {\footnotesize{}$-0.533\pm0.010$}\tabularnewline
\hline 
{\footnotesize{}90,-45 } & {\footnotesize{}873} & {\footnotesize{}2686} & {\footnotesize{}2644} & {\footnotesize{}730} & {\footnotesize{}6933} & {\footnotesize{}$-0.538\pm0.010$}\tabularnewline
\hline 
{\footnotesize{}90,\hphantom{-}45 } & {\footnotesize{}2696} & {\footnotesize{}966} & {\footnotesize{}902} & {\footnotesize{}2453} & {\footnotesize{}7017} & {\footnotesize{}$\phantom{-}0.468\pm0.011$}\tabularnewline
\hline 
\hline 
{\footnotesize{}$S$} & \multicolumn{4}{c||}{} & {\footnotesize{}27885} & {\footnotesize{}$\phantom{-}\mathbf{2.085\pm0.020}$}\tabularnewline
\hline 
\hline 
{\footnotesize{}$P_{m}$} & \multicolumn{5}{c||}{} & {\footnotesize{}$6.448\cdot10^{-4}$}\tabularnewline
\hline 
{\footnotesize{}$P_{g}$} & \multicolumn{5}{c||}{{\footnotesize{}21207 wins}} & {\footnotesize{}$6.538\cdot10^{-5}$}\tabularnewline
\hline 
\end{tabular}{\footnotesize{}\hfill{}}%
\begin{tabular}{|c||c|c|c|c||c||c|}
\hline 
\multicolumn{7}{|c|}{{\footnotesize{}Nov. 11, 2015 - June 24, 2016,$\Psi^{-}$}}\tabularnewline
\hline 
\hline 
{\footnotesize{}$\alpha,\beta$} & {\footnotesize{}$\uparrow\uparrow$} & {\footnotesize{}$\uparrow\downarrow$} & {\footnotesize{}$\downarrow\uparrow$} & {\footnotesize{}$\downarrow\downarrow$} & {\footnotesize{}total } & {\footnotesize{}$\left\langle \sigma_{\alpha}\sigma_{\beta}\right\rangle $}\tabularnewline
\hline 
\hline 
{\footnotesize{}0,-45 } & {\footnotesize{}817} & {\footnotesize{}2596} & {\footnotesize{}2873} & {\footnotesize{}742} & {\footnotesize{}7028} & {\footnotesize{}$-0.556\pm0.010$}\tabularnewline
\hline 
{\footnotesize{}0,\hphantom{-}45 } & {\footnotesize{}696} & {\footnotesize{}2570} & {\footnotesize{}2788} & {\footnotesize{}772} & {\footnotesize{}6826} & {\footnotesize{}$-0.570\pm0.010$}\tabularnewline
\hline 
{\footnotesize{}90,-45 } & {\footnotesize{}2783} & {\footnotesize{}787} & {\footnotesize{}840} & {\footnotesize{}2503} & {\footnotesize{}6913} & {\footnotesize{}$\phantom{-}0.529\pm0.010$}\tabularnewline
\hline 
{\footnotesize{}90,\hphantom{-}45 } & {\footnotesize{}865} & {\footnotesize{}2620} & {\footnotesize{}2640} & {\footnotesize{}791} & {\footnotesize{}6916} & {\footnotesize{}$-0.521\pm0.010$}\tabularnewline
\hline 
\hline 
{\footnotesize{}$S$} & \multicolumn{4}{c||}{} & {\footnotesize{}27683} & {\footnotesize{}$\mathbf{2.177\pm0.020}$}\tabularnewline
\hline 
\hline 
{\footnotesize{}$P_{m}$} & \multicolumn{5}{c||}{} & {\footnotesize{}$8.932\cdot10^{-16}$}\tabularnewline
\hline 
{\footnotesize{}$P_{g}$} & \multicolumn{5}{c||}{{\footnotesize{}21373 wins}} & {\footnotesize{}$4.527\cdot10^{-17}$}\tabularnewline
\hline 
\end{tabular}
\par\end{centering}{\footnotesize \par}
\caption{Combined experimental data.}
\end{table}

\begin{center}
\begin{table}[h]
\begin{centering}
\begin{tabular}{|c|c|}
\hline 
method & S-value\tabularnewline
\hline 
\hline 
weighted arithmetic mean & $2.131\pm0.0141$\tabularnewline
\hline 
event based & $2.130\pm0.0144$\tabularnewline
\hline 
\end{tabular}
\par\end{centering}
\medskip{}

\begin{centering}
\begin{tabular}{|c|c|}
\hline 
method & P-value\tabularnewline
\hline 
\hline 
$P_{m}$ with combined S-value & $1.017\cdot10^{-16}$\tabularnewline
\hline 
$P_{g}$ with $42580$ wins of $55568$ & $4.891\cdot10^{-18}$\tabularnewline
\hline 
\end{tabular}
\par\end{centering}
\caption{Evaluation of combined $S$- and P-values for the complete dataset.}
\end{table}
\par\end{center}

\newpage{}

\section{Independence of Random Bits and No-Signaling}

The space-like separation of the measurements in the experiment also
implies independence of random bits generated in the two labs. Furthermore
there should be no correlations between local outcomes and distant
measurement settings, as this would require superluminal communication
(signaling). As was pointed out in \cite{Bednorz2015,Adenier2015},
this should be tested as it would (at least) indicate experimental
problems possibly disvalidating the Bell test. We thus check our experimental
data for correlations between random bits from lab~1 and lab~2 and
whether the random input at lab~1 is correlated with the measurement
outcome of lab~2 or vice versa.

\subsection{Independence of random bits}

\begin{table}[h]
\centering{} %
\begin{tabular}{cc}
Nov. 27, 2015 & April 7, 2016\tabularnewline
\begin{tabular}{c|c|c|c}
 & $b=0$ & $b=1$ & \tabularnewline
\hline 
$a=0$ & 79 & 69 & $\phantom{0}$148\tabularnewline
\hline 
$a=1$ & 78 & 74 & $\phantom{0}$152\tabularnewline
\hline 
 & $\phantom{0}$157 & $\phantom{0}$143 & \tabularnewline
\end{tabular} & %
\begin{tabular}{c|c|c|c}
 & $b=0$ & $b=1$ & \tabularnewline
\hline 
$a=0$ & 955 & 908 & 1863\tabularnewline
\hline 
$a=1$ & 912 & 936 & 1648\tabularnewline
\hline 
 & 1867 & 1844 & \tabularnewline
\end{tabular}\tabularnewline
$P=0.72$ & $P=0.24$\tabularnewline
Apr. 15, 2016 & June 14, 2016\tabularnewline
\begin{tabular}{c|c|c|c}
 & $b=0$ & $b=1$ & \tabularnewline
\hline 
$a=0$ & 2539 & 2461 & 5000\tabularnewline
\hline 
$a=1$ & 2502 & 2498 & 5000\tabularnewline
\hline 
 & 5041 & 4959 & \tabularnewline
\end{tabular} & %
\begin{tabular}{c|c|c|c}
 & $b=0$ & $b=1$ & \tabularnewline
\hline 
$a=0$ & 2565 & 2506 & 5071\tabularnewline
\hline 
$a=1$ & 2406 & 2523 & 4929\tabularnewline
\hline 
 & 4971 & 5029 & \tabularnewline
\end{tabular}\tabularnewline
$P=0.45$ & $P=0.08$\tabularnewline
\end{tabular}\caption{\label{tab:Independence-RN}Random bits used for selection of measurement
settings in presented measurement runs.}
\end{table}

For independent random bits neither the probability for $b=1$ or
$b=0$ should depend on the space-like separated selection of random
variable $a$ nor the probability of $a$ on $b$. To test this hypothesis
we perform a two-proportion z-test, since the distribution of the
random bits is binomial and thus for large $N$ approaches Gaussian
with a known standard deviation. The calculated P-values for our data
give no reason to discard the null-hypothesis of independent random
bits.

\subsection{No-signaling}

Next, we check if the local measurements are correlated with the random
inputs on the other side of the experiment. Since we have no sufficiently
exact prediction of the outcome probabilities (due to the details
of the measurement procedure and available statistics), we employ
a two-sample t-test to check the null hypothesis that the measurement
outcomes do not depend on the random input of the other side. The
distribution of the calculated P-values (Table \ref{tab:no-signaling})
provides no evidence to reject the no-signaling assumption.

\begin{table}[h]
\centering{} %
\begin{tabular}{cc}
\multicolumn{2}{c}{Nov. 27, 2015}\tabularnewline
\begin{tabular}{c|c|c|c}
 & $y=+1$ & $y=-1$ & \tabularnewline
\hline 
$a=0$ & 80 & 68 & $\phantom{0}$148\tabularnewline
\hline 
$a=1$ & 76 & 76 & $\phantom{0}$152\tabularnewline
\hline 
 & $\phantom{0}$156 & $\phantom{0}$144 & \tabularnewline
\end{tabular} & %
\begin{tabular}{c|c|c|c}
 & $x=+1$ & $x=-1$ & \tabularnewline
\hline 
$b=0$ & 88 & 69 & $\phantom{0}$157\tabularnewline
\hline 
$b=1$ & 73 & 70 & $\phantom{0}$143\tabularnewline
\hline 
 & $\phantom{0}$161 & $\phantom{0}$139 & \tabularnewline
\end{tabular}\tabularnewline
$P=0.483$ & $P=0.387$\tabularnewline
\multicolumn{2}{c}{Apr. 07, 2016}\tabularnewline
\begin{tabular}{c|c|c|c}
 & $y=+1$ & $y=-1$ & \tabularnewline
\hline 
$a=0$ & 922 & 941 & 1863\tabularnewline
\hline 
$a=1$ & 989 & 859 & 1848\tabularnewline
\hline 
 & 1911 & 1800 & \tabularnewline
\end{tabular} & %
\begin{tabular}{c|c|c|c}
 & $x=+1$ & $x=-1$ & \tabularnewline
\hline 
$b=0$ & 966 & 901 & 1867\tabularnewline
\hline 
$b=1$ & 950 & 894 & 1844\tabularnewline
\hline 
 & 1916 & 1795 & \tabularnewline
\end{tabular}\tabularnewline
$P=0.014$ & $P=0.892$\tabularnewline
\multicolumn{2}{c}{Apr. 15, 2016}\tabularnewline
\begin{tabular}{c|c|c|c}
 & $y=+1$ & $y=-1$ & \tabularnewline
\hline 
$a=0$ & 2603 & 2397 & 5000\tabularnewline
\hline 
$a=1$ & 2636 & 2364 & 5000\tabularnewline
\hline 
 & 5239 & 4761 & \tabularnewline
\end{tabular} & %
\begin{tabular}{c|c|c|c}
 & $x=+1$ & $x=-1$ & \tabularnewline
\hline 
$b=0$ & 2531 & 2510 & 5041\tabularnewline
\hline 
$b=1$ & 2480 & 2479 & 4959\tabularnewline
\hline 
 & 5011 & 4989 & \tabularnewline
\end{tabular}\tabularnewline
$P=0.509$ & $P=0.890$\tabularnewline
\multicolumn{2}{c}{June 14, 2016}\tabularnewline
\begin{tabular}{c|c|c|c}
 & $y=+1$ & $y=-1$ & \tabularnewline
\hline 
$a=0$ & 2464 & 2607 & 5071\tabularnewline
\hline 
$a=1$ & 2400 & 2529 & 4929\tabularnewline
\hline 
 & 4864 & 5138 & \tabularnewline
\end{tabular} & %
\begin{tabular}{c|c|c|c}
 & $x=+1$ & $x=-1$ & \tabularnewline
\hline 
$b=0$ & 2459 & 2512 & 4971\tabularnewline
\hline 
$b=1$ & 2545 & 2484 & 5029\tabularnewline
\hline 
 & 5004 & 4996 & \tabularnewline
\end{tabular}\tabularnewline
$P=0.919$ & $P=0.255$\tabularnewline
\end{tabular}\caption{\label{tab:no-signaling}Data for testing no-signaling.}
\end{table}

\end{document}